# Casimir energy for $N$ constant conductivity $\delta$-plates with a neural network perception

Venkat Abhignan
*Qdit Labs Pvt. Ltd., Bengaluru - 560092, India*

The Casimir energy for $N$ $\delta$-function plates depends on the multiple scattering parameter $\Delta$. This $N$-body interaction was distributed into interactions with nearest neighbour scattering and next-to-nearest neighbour scattering based on partitions of $N-1$ and its permutations. Implementing this methodology, we investigate the Casimir interaction for multiple plates with constant conductivity relatable to Graphene. We also study the Casimir energy between a perfect magnetic conductor and multiple constant conductivity $\delta$ plates, which results in Boyer repulsion. In the asymptotic limit for ideal boundary conditions, the results become simple where the multiple scattering parameter $\Delta$ consists only of the nearest neighbour scattering term. Further, we used neural networks to analyze the Casimir energy in the Boyer repulsion configurations to understand the influence of pairwise energies on the many-body energy. The neural network could distinguish the repulsive and attractive forces depending on the regions of varying conductivity.

## I. INTRODUCTION

Casimir showed that there is a force between two parallel, perfectly conducting plates when there are quantum electrodynamic fluctuations in a vacuum [1]. Consequently, it was shown for two dielectric slabs with finite conductivity [2]. It is acknowledged that the non-additivity of this interaction makes it difficult to calculate Casimir energies and related Lifshitz forces for multiple dielectric bodies [3, 4], even if exact results have been obtained for ideal boundary conditions.

The Casimir force may significantly contribute to stiction and the adherence of microscopic components, and hence may challenge the functioning of nano- or micro-structured systems. However, the Casimir force can be utilized to make a new form of microelectromechanical systems [5, 6]. Casimir energies can be expressed using the multiple scattering formalism in terms of the reflection coefficients of the structure [7–10]. A single layer of Graphite is called Graphene, where carbon atoms align to create a two-dimensional hexagonal lattice [11]. Graphene, with its miniaturized structure and unique optical properties [12] can be utilized to modulate Casimir forces by modifying reflection coefficients [13–16]. The advancements in micro- and nano-electromechanical systems using Graphene [17] have made the study in its Casimir force technologically significant [18, 19].

Tomaš initially pursued the Casimir force for parallel multilayered structures [20, 21]. Casimir energy for $N$ layers of Graphene with optical Fresnel coefficients had been handled recently using this approach [22]. The initial study for $N$ layers of Graphene considered a scalar model and was solved using the mode-summation approach with zeta function regularisation [23]. Other than these formalisms, the Piston approach [24], the Modal approach [25], the path-integral approach [26] and the plasma sheets model [27] have been applied to study Casimir energy for multilayers. However, no closed-form Casimir calculation for $N$ bodies exists. Adding more layers results in complex expressions and difficulty when dealing with multiple dielectric cavities. Beyond the recent technological application of many-body Casimir physics in nanocomposites [28] nanophotonics [29] and nanotechnology [30], it is of fundamental interest as well. The Casimir force between three bodies is crucial when investigating gravity at short distances [31–33]. A novel force beyond the standard model has been sought after by applying many-body Casimir measurements with high precision [34].

Casimir energy of $N$ magnetodielectric $\delta$-function plates was predicted recently using multiple scattering formalism and stress-tensor method [35, 36]. $\delta$-function plates [37] and related plasma sheet model [38, 39] have been developed exclusively to study 2D materials like Graphene. In this work, we calculate the Casimir energy of Graphene multilayers using $\delta$-function plates with constant and isotropic optical conductivity [40]. The optical response of Graphene, which is essential for Casimir calculations, is determined by its electronic structure and Dirac-like carriers [15, 41]. The optical conductivity is independent of all material parameters in the limit for frequencies $\hbar\omega \leq 3eV$ [42]. The optical absorption is a meager 2.3% at room temperature, depending only on the fine structure constant $\alpha = 1/137$ [43]. Also, if spatial dispersion can be ignored, the conductivity is isotropic [44].

The promise of neural networks for research in fundamental physics has drawn more and more attention [45, 46]. When collecting and evaluating information to create models that can predict the behavior of complicated frameworks like quantum many-body systems, neural networks are helpful. Casimir energy under ideal boundary conditions was recently handled in a scalar model using neural networks to learn and predict the energy by varying the boundaries [47].

In Sec.II, we describe the method to obtain a generalized expression for the Casimir energy between $N$ $\delta$-plates. We also consider an instance to get a closed-form expression of the Casimir energy and the corresponding force for $N=6$ plates with different properties. In Sec. III, we calculate and analyze the Casimir energy of $N=2,\cdots,6$ equidistant plates with constant conductivity relevant to Graphene. In Sec. IV, we investigate the repulsive Casimir forces resulting in an interaction between infinitely permeable material and $\delta$ plates with constant conductivity. Further, we were interested in analyzing the many-body Casimir energy using neural networks, compared to the pairwise energies, to gain insights into repulsive and attractive Casimir forces. In Sec. IVA, we described the neural network architecture used, the role of backpropagation used for weight optimization, and how the learned weights provided insights.

## II. CASIMIR ENERGY AND FORCE OF $N$ MAGNETODIELECTRIC $\delta$-PLATES

The multiple scattering parameter $\Delta_{12\cdots N}$ can describe the Casimir energy $\Delta E_{(12\cdots N)}$ of $N$ plates such as [35]

$$\frac{\Delta E_{(12\cdots N)}}{A} = \frac{1}{2}\int_{-\infty}^{\infty}\frac{d\zeta}{2\pi}\int\frac{d^2k_\perp}{(2\pi)^2}\left[\ln\left[\Delta^H_{12\cdots N}\right]+\ln\left[\Delta^E_{12\cdots N}\right]\right], \tag{1}$$

integrated over all wavenumbers and frequencies (with superscript $E$ denoting TE mode and $H$ for TM mode).

The parameter $\Delta_{12\cdots N}$ can be distributed into the nearest neighbour scattering parameter $\Delta_{ij}$, and the next-to-nearest neighbour, next-to-next-to-nearest neighbour, $\cdots$ scattering parameters $\Delta_{ik}$ based on partitions of $N-1$ and its permutations (Further, described with an example for $N=6$ in Eq. (6)). The characteristics of scattering parameters indicate the various ways the propagation may contribute to the energy. $\Delta_{ij}$ and $\Delta_{ik}$ can be produced by visualizing a diagrammatic loop distribution with an exponential dependence on the distance between the plates $l_{ij}$, and in terms of the optical properties of the plates with reflection and transmission coefficients $r_i$ and $t_i$ such as,

$$r_i^H = -\frac{\lambda_{gi}^\perp\zeta^2}{\lambda_{gi}^\perp\zeta^2+2\kappa}+\frac{\lambda_{ei}^\perp\kappa}{\lambda_{ei}^\perp\kappa+2},\ t_i^H = 1-\frac{\lambda_{gi}^\perp\zeta^2}{\lambda_{gi}^\perp\zeta^2+2\kappa}-\frac{\lambda_{ei}^\perp\kappa}{\lambda_{ei}^\perp\kappa+2} \tag{2}$$

with $\kappa=\sqrt{k_\perp^2+\zeta^2}$. Coefficients $r^E, t^E$ corresponding to TE mode can be obtained by replacing superscripts $H\rightarrow E$ and by swapping $\lambda_{ei}^\perp\leftrightarrow\lambda_{gi}^\perp$. The matrix

$$\boldsymbol{\lambda_{e,g}}(\zeta) = \begin{bmatrix}\lambda_{e,g}^\perp(\zeta) & 0 & 0\\ 0 & \lambda_{e,g}^\perp(\zeta) & 0\\ 0 & 0 & 0\end{bmatrix} \tag{3}$$

describes the electric properties $\boldsymbol{\lambda_e}$ and magnetic properties $\boldsymbol{\lambda_g}$ corresponding to $\boldsymbol{\varepsilon}$ and $\boldsymbol{\mu}$ of the material, respectively (with planar symmetry, implying isotropic and homogeneous on the plate) in Heaviside–Lorentz units. Based on these coefficients, the nearest neighbour scattering parameters $\Delta_{ij}$ are

$$\Delta_{ij} = 1 - r_i e^{-\kappa l_{ij}} r_j e^{-\kappa l_{ij}}, \tag{4}$$

for $j=i+1$ ($i\in[1,N-1]$ where $i$ and $j$ are adjacent plates) and next-to-next-to-nearest neighbour, $\cdots$ scattering parameters $\Delta_{ik}$ are

$$\Delta_{ik} = -r_i e^{-\kappa l_{i,i+1}} t_{i+1} e^{-\kappa l_{i+1,i+2}} t_{i+2}\cdots e^{-\kappa l_{k-1,k}} r_k e^{-\kappa l_{k-1,k}}\cdots t_{i+1} e^{-\kappa l_{i,i+1}}, \tag{5}$$

for $k\geq i+2$ ($i\in[1,N-2]$ where $i$ and $k$ are not adjacent plates). $\Delta_{ij}$ only depends on the reflection coefficients of neighbouring plates $i,j$ with exponential dependence of length between the plates. $\Delta_{ik}$ depends on the reflection coefficients of bordering plates $i,k$ and transmission coefficients of nearby adjacent plates $i+1,\cdots,k-1$ between the bordering plates as the propagation happens with exponential length dependence.

For example, the Casimir energy of a $N=6$ plate configuration, which is usually hard to handle, depends only on the multiple scattering parameter $\Delta_{123456}$. This term can be separated based on partitions of $N-1=5$ as

$$\begin{aligned}\Delta_{123456} &= \Delta_{12}\Delta_{23}\Delta_{34}\Delta_{45}\Delta_{56}+\Delta_{12}\Delta_{24}\Delta_{45}\Delta_{56}+\Delta_{13}\Delta_{34}\Delta_{45}\Delta_{56}+\Delta_{12}\Delta_{23}\Delta_{35}\Delta_{56}+\Delta_{12}\Delta_{23}\Delta_{34}\Delta_{46}+\Delta_{13}\Delta_{35}\Delta_{56}\\ &+\Delta_{13}\Delta_{34}\Delta_{46}+\Delta_{12}\Delta_{24}\Delta_{46}+\Delta_{12}\Delta_{23}\Delta_{36}+\Delta_{12}\Delta_{25}\Delta_{56}+\Delta_{14}\Delta_{45}\Delta_{56}+\Delta_{14}\Delta_{46}+\Delta_{13}\Delta_{36}+\Delta_{15}\Delta_{56}+\Delta_{12}\Delta_{26}+\Delta_{16}\end{aligned} \tag{6}$$

and Fig. 1 illustrates the diagrammatic loop distribution of this multiple scattering parameter. Understanding the propagation of the multiple scattering formalism is easier with the help of this pattern. The partitions of 5 are (5), (4,1), (3,2), (3,1,1), (2,2,1), (2,1,1,1), and (1,1,1,1,1). In Eq.(6), it can be intuitively understood that term corresponding to partition (5) is $\Delta_{16}$ and partition (1,1,1,1,1) is $\Delta_{12}\Delta_{23}\Delta_{34}\Delta_{45}\Delta_{56}$. Similarly, terms corresponding to partition (4,1), (1,4) are $\Delta_{15}\Delta_{56}, \Delta_{12}\Delta_{26}$ and partition (3,2), (2,3) are $\Delta_{14}\Delta_{46}, \Delta_{13}\Delta_{36}$, respectively. Terms corresponding to partition (1,1,3), (1,3,1), (3,1,1) are $\Delta_{12}\Delta_{23}\Delta_{36}, \Delta_{12}\Delta_{25}\Delta_{56}, \Delta_{14}\Delta_{45}\Delta_{56}$, respectively. Terms corresponding to partition (2,2,1), (2,1,2), (1,2,2) are $\Delta_{13}\Delta_{35}\Delta_{56}, \Delta_{13}\Delta_{34}\Delta_{46}, \Delta_{12}\Delta_{24}\Delta_{46}$, respectively. And terms corresponding to partition (1,2,1,1), (2,1,1,1), (1,1,2,1), (1,1,1,2) are $\Delta_{12}\Delta_{24}\Delta_{45}\Delta_{56}, \Delta_{13}\Delta_{34}\Delta_{45}\Delta_{56}, \Delta_{12}\Delta_{23}\Delta_{35}\Delta_{56}, \Delta_{12}\Delta_{23}\Delta_{34}\Delta_{46}$, respectively. Further, using Eqs. (4) and (5) in this multiple scattering parameter Eq. (6) for $N = 6$ plates in terms of optical properties in Eq. (2) gives

$$
\begin{aligned}
\Delta_{123456} = &\left(1 - r_1 r_2 e^{-2\kappa l_{12}}\right)\left(1 - r_2 r_3 e^{-2\kappa l_{23}}\right)\left(1 - r_3 r_4 e^{-2\kappa l_{34}}\right)\left(1 - r_4 r_5 e^{-2\kappa l_{45}}\right)\left(1 - r_5 r_6 e^{-2\kappa l_{56}}\right)\\
&- \left(1 - r_1 r_2 e^{-2\kappa l_{12}}\right) r_2 t_3^2 r_4 e^{-2\kappa(l_{23}+l_{34})}\left(1 - r_4 r_5 e^{-2\kappa l_{45}}\right)\left(1 - r_5 r_6 e^{-2\kappa l_{56}}\right)\\
&- r_1 t_2^2 r_3 e^{-2\kappa(l_{12}+l_{23})}\left(1 - r_3 r_4 e^{-2\kappa l_{34}}\right)\left(1 - r_4 r_5 e^{-2\kappa l_{45}}\right)\left(1 - r_5 r_6 e^{-2\kappa l_{56}}\right)\\
&- \left(1 - r_1 r_2 e^{-2\kappa l_{12}}\right)\left(1 - r_2 r_3 e^{-2\kappa l_{23}}\right) r_3 t_4^2 r_5 e^{-2\kappa(l_{34}+l_{45})}\left(1 - r_5 r_6 e^{-2\kappa l_{56}}\right)\\
&- \left(1 - r_1 r_2 e^{-2\kappa l_{12}}\right)\left(1 - r_2 r_3 e^{-2\kappa l_{23}}\right)\left(1 - r_3 r_4 e^{-2\kappa l_{34}}\right) r_4 t_5^2 r_6 e^{-2\kappa(l_{45}+l_{56})}\\
&+ r_1 t_2^2 r_3^2 t_4^2 r_5 e^{-2\kappa(l_{12}+l_{23}+l_{34}+l_{45})}\left(1 - r_5 r_6 e^{-2\kappa l_{56}}\right) + r_1 t_2^2 r_3\left(1 - r_3 r_4 e^{-2\kappa l_{34}}\right) r_4 t_5^2 r_6 e^{-2\kappa(l_{12}+l_{23}+l_{45}+l_{56})}\\
&+ \left(1 - r_1 r_2 e^{-2\kappa l_{12}}\right) r_2 t_3^2 r_4^2 t_5^2 r_6 e^{-2\kappa(l_{23}+l_{34}+l_{45}+l_{56})} - \left(1 - r_1 r_2 e^{-2\kappa l_{12}}\right)\left(1 - r_2 r_3 e^{-2\kappa l_{23}}\right) r_3 t_4^2 t_5^2 r_6 e^{-2\kappa(l_{34}+l_{45}+l_{56})}\\
&- \left(1 - r_1 r_2 e^{-2\kappa l_{12}}\right) r_2 t_3^2 t_4^2 r_5 e^{-2\kappa(l_{23}+l_{34}+l_{45})}\left(1 - r_5 r_6 e^{-2\kappa l_{56}}\right) - r_1 t_2^2 t_3^2 r_4 e^{-2\kappa(l_{12}+l_{23}+l_{34})}\left(1 - r_4 r_5 e^{-2\kappa l_{45}}\right)\left(1 - r_5 r_6 e^{-2\kappa l_{56}}\right)\\
&+ r_1 t_2^2 t_3^2 r_4^2 t_5^2 r_6 e^{-2\kappa(l_{12}+l_{23}+l_{34}+l_{45}+l_{56})} + r_1 t_2^2 r_3^2 t_4^2 t_5^2 r_6 e^{-2\kappa(l_{12}+l_{23}+l_{34}+l_{45}+l_{56})} - r_1 t_2^2 t_3^2 t_4^2 r_5 e^{-2\kappa(l_{12}+l_{23}+l_{34}+l_{45})}\left(1 - r_5 r_6 e^{-2\kappa l_{56}}\right)\\
&- \left(1 - r_1 r_2 e^{-2\kappa l_{12}}\right) r_2 t_3^2 t_4^2 t_5^2 r_6 e^{-2\kappa(l_{23}+l_{34}+l_{45}+l_{56})} - r_1 t_2^2 t_3^2 t_4^2 t_5^2 r_6 e^{-2\kappa(l_{12}+l_{23}+l_{34}+l_{45}+l_{56})}
\end{aligned}
\tag{7}
$$

with length $l_{ij}$ between the adjacent plates $i$ and $j$.

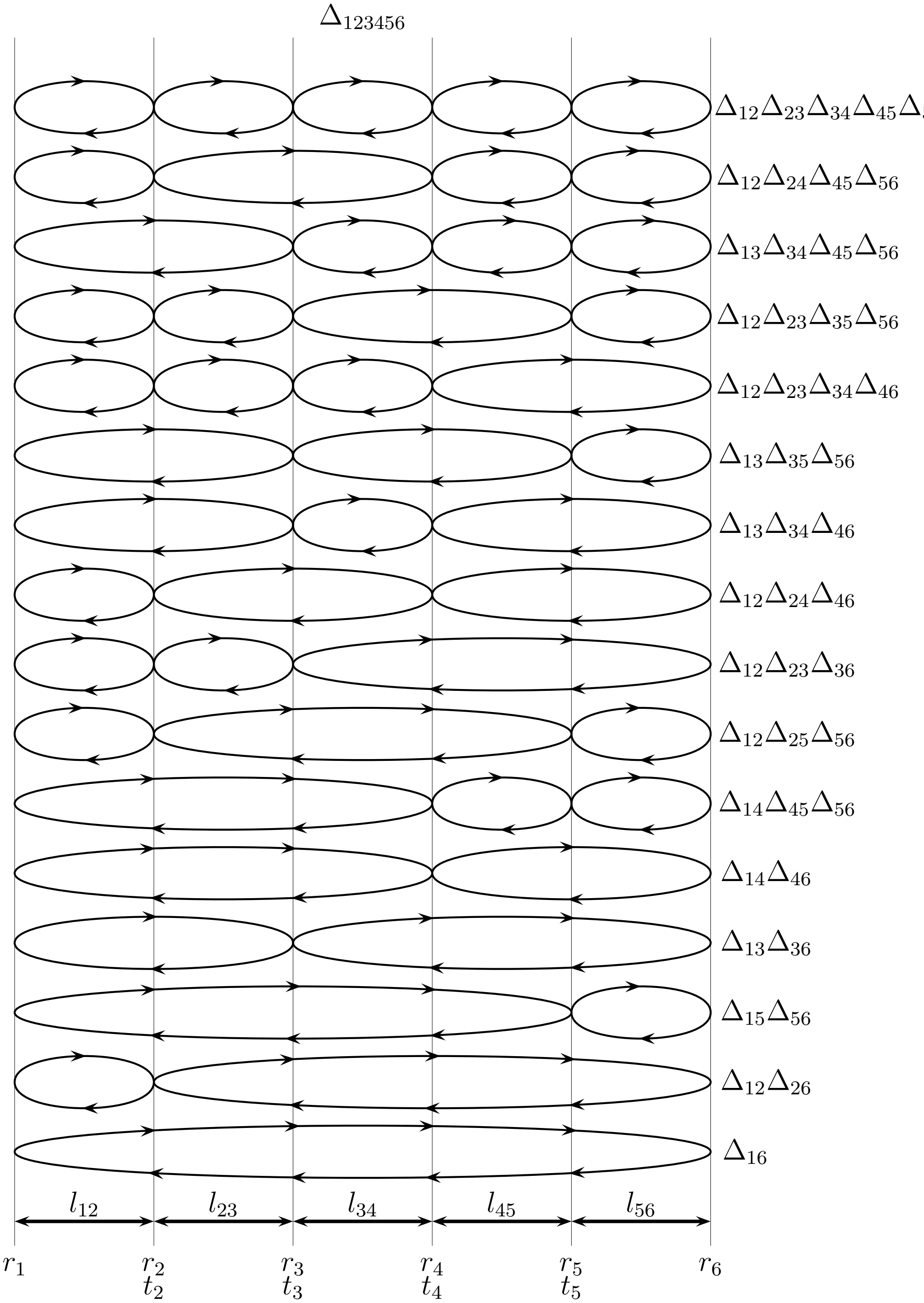

FIG. 1: Multiple scattering parameter $\Delta_{123456}$ is illustrated by partitions consisting of nearest neighbour scattering and next-to-nearest neighbour scattering. Each scattering term is defined in a tractable manner from the visualization of the loop based on the optical properties of the plates denoted by $r_i$ and $t_i$, the reflection and transmission coefficients, respectively, along with the distance between the plates where $i$ and $j$ are adjacent is denoted by $l_{ij}$. For instance, term $\Delta_{36} = r_3 e^{-\kappa l_{34}} t_4 e^{-\kappa l_{45}} t_5 e^{-\kappa l_{56}} r_6 e^{-\kappa l_{56}} t_5 e^{-\kappa l_{45}} t_4 e^{-\kappa l_{34}}$ from Eq. (5) refers to loop between plates $i = 3$ and $k = 6$ with initial reflection $r_3$, propagation with exponential dependence of length $l_{34}$, transmission $t_4$, propagation of length $l_{45}$, transmission $t_5$, propagation of length $l_{56}$, reflection with $r_6$, again propagation of length $l_{56}$, transmission with $t_5$, propagation of length $l_{45}$, transmission $t_4$, propagation of length $l_{34}$ and it continues.

The quantum vacuum's physical manifestation is the corresponding Casimir force. Casimir force $\mathbf{F_i}$ per unit area or the pressure $P_i$ on the plate $i$ can be related to the Casimir energy $\Delta E_{(123456)}$ of $N = 6$ plates using Eq. (7) in Eq.(1) as

$$
P_i = \frac{\mathbf{F_i}\cdot\hat{\mathbf{z}}}{A} = \begin{cases} \frac{1}{A}\frac{\partial \Delta E_{(1\cdots 6)}}{\partial l_{i,i+1}}, & i = 1\\ \frac{1}{A}\frac{\partial \Delta E_{(1\cdots 6)}}{\partial l_{i,i+1}} - \frac{1}{A}\frac{\partial \Delta E_{(1\cdots 6)}}{\partial l_{i-1,i}}, & i = 2, \cdots, 5\\ -\frac{1}{A}\frac{\partial \Delta E_{(1\cdots 6)}}{\partial l_{i-1,i}}, & i = 6. \end{cases}
\tag{8}
$$

This expression can be better understood for a simple case of $N = 3$ plates at positions $z = p_i, i = 1, 2, 3$ where energy $\Delta E_{(123)}$ is a function of $(p_2 - p_1 = l_{12}, p_3 - p_2 = l_{23})$, the relative positions of the plates and the force on the plate $i$ becomes

$$
P_i = \frac{\mathbf{F_i}\cdot\hat{\mathbf{z}}}{A} = -\frac{1}{A}\frac{\partial \Delta E_{(123)}}{\partial p_i} = \begin{cases} \frac{1}{A}\frac{\partial \Delta E_{(123)}}{\partial l_{12}}, & i = 1\\ -\frac{1}{A}\frac{\partial \Delta E_{(123)}}{\partial l_{12}} + \frac{1}{A}\frac{\partial \Delta E_{(123)}}{\partial l_{23}}, & i = 2\\ -\frac{1}{A}\frac{\partial \Delta E_{(123)}}{\partial l_{23}}, & i = 3. \end{cases}
\tag{9}
$$

## III. CASIMIR ENERGY BETWEEN CONSTANT CONDUCTIVITY $\delta$-PLATES

Based on massless Dirac model [15, 41] for constant isotropic conductivity $\sigma$,

$$
\lambda_e^{\perp}(\zeta) = \frac{\sigma}{\zeta}
\tag{10}
$$

gives the dielectric optical property in Eq. (3) for $\delta$-plate. With this, the reflection and transmission coefficients ($r_\sigma$ and $t_\sigma$, with $\sigma$ in subscript denoting the $\delta$-plates with $\sigma$ conductivity) from Eq. (2) are

$$
r_\sigma^H = \frac{\sigma\kappa}{\sigma\kappa + 2\zeta},\; r_\sigma^E = -\frac{\sigma\zeta}{\sigma\zeta + 2\kappa} \text{ and } t_\sigma^H = 1 - \frac{\sigma\kappa}{\sigma\kappa + 2\zeta},\; t_\sigma^E = 1 - \frac{\sigma\zeta}{\sigma\zeta + 2\kappa},
\tag{11}
$$

with no magnetic property $\lambda_g^{\perp} = 0$. Using this the Casimir energy of $N = 2$ plates with constant conductivity from Eq. (1) is

$$
\frac{\Delta E_{(12)}}{A} = \frac{1}{2}\int_{-\infty}^{\infty}\frac{d\zeta}{2\pi}\int\frac{d^2k_\perp}{(2\pi)^2}\left[\ln\left[1 - \left(\frac{\sigma\kappa}{\sigma\kappa + 2\zeta}\right)^2 e^{-2\kappa a}\right] + \ln\left[1 - \left(-\frac{\sigma\zeta}{\sigma\zeta + 2\kappa}\right)^2 e^{-2\kappa a}\right]\right]
\tag{12}
$$

from considering $r_{1,2} = r_\sigma$, $r_{3,4,5,6} = 0$ and $l_{ij} = a$ in Eq. (7). Taking the asymptotic limit $\sigma \to \infty$ we obtain the Casimir energy between two perfectly conducting plates as $\Delta E^c_{(12)}/A = -\pi^2/720a^3$. This leads to an attractive force between the two plates since $\mathbf{F_1} \cdot \hat{\mathbf{z}}/A = \pi^2/240a^4$ and $\mathbf{F_2} \cdot \hat{\mathbf{z}}/A = -\pi^2/240a^4$ from Eq. (8). Introducing spherical polar coordinates $k_\perp = \kappa \sin\theta$ and $\zeta = \kappa \cos\theta$ [48] in Eq. (12) we obtain

$$\frac{\Delta E_{(12)}}{\Delta E^c_{(12)}} = -\frac{45}{2\pi^4}\int_0^1 dt \int_0^\infty s^2 ds \left[\ln\left[1-\left(\frac{\sigma}{\sigma+2t}\right)^2 e^{-s}\right] + \ln\left[1-\left(-\frac{\sigma}{\sigma+\frac{2}{t}}\right)^2 e^{-s}\right]\right] \tag{13}$$

scaled with $\Delta E^c_{(12)}$. Evaluation of $s$ integral yields

$$\frac{\Delta E_{(12)}}{\Delta E^c_{(12)}} = \frac{45}{\pi^4}\int_0^1 dt \left[\mathrm{Li}_4\left(\frac{\sigma}{\sigma+2t}\right)^2 + \mathrm{Li}_4\left(-\frac{\sigma}{\sigma+\frac{2}{t}}\right)^2\right] \tag{14}$$

where $\mathrm{Li}_4(z)$ is a polylogarithm function and numerical evaluation of Eq. (14) considering $\sigma = \pi\alpha$ for Graphene gives $\Delta E_{(12)} = 0.00538\Delta E^c_{(12)}$ as shown previously in Ref. [16].

Explicit details for calculating the Casimir energy for $N = 3, 4$ $\delta$-plates with $\sigma$ conductivity are given in the Appendix. The numerical evaluation using $\sigma = \pi\alpha$ yields Casimir energy for $N = 3, 4$ Graphene plates as $\Delta E_{(123)} = 0.011\Delta E^c_{(12)}$, $\Delta E_{(1234)} = 0.017\Delta E^c_{(12)}$. In a similar manner we evaluate Casimir energy of $N = 5, 6$ Graphene plates as $\Delta E_{(12345)} = 0.022\Delta E^c_{(12)}$, $\Delta E_{(123456)} = 0.028\Delta E^c_{(12)}$. In Fig. 2, the energy of $N = 2, 3, 4, 5, 6$ plates scaled by $\Delta E^c_{(12)}$ and normalised per unit plate $N$ is displayed to understand how the energy behaves for different values of $\sigma$. The energy $\Delta E_{(12\cdots N)}/N$ approaches the $\Delta E^c_{(12)}$ for two perfectly conducting plates as $N >> 1$ which was previously seen in Ref. [23]. This is in the case of ideal boundary conditions like perfectly conducting plates (Dirichlet boundary conditions [49]) because the Casimir energy of $N$ plates in Eq. (1) only involves the nearest neighbour scattering parameters $\Delta_{ij}$ in Eq. (4). The analysis is simple when the next-to-next-to-nearest neighbour, $\cdots$ scattering parameters $\Delta_{ik}$ in Eq. (5) do not contribute, and it can be evaluated in a straightforward manner that

$$\Delta E^c_{(12\cdots N)} = (N-1)\Delta E^c_{(12)} \tag{15}$$

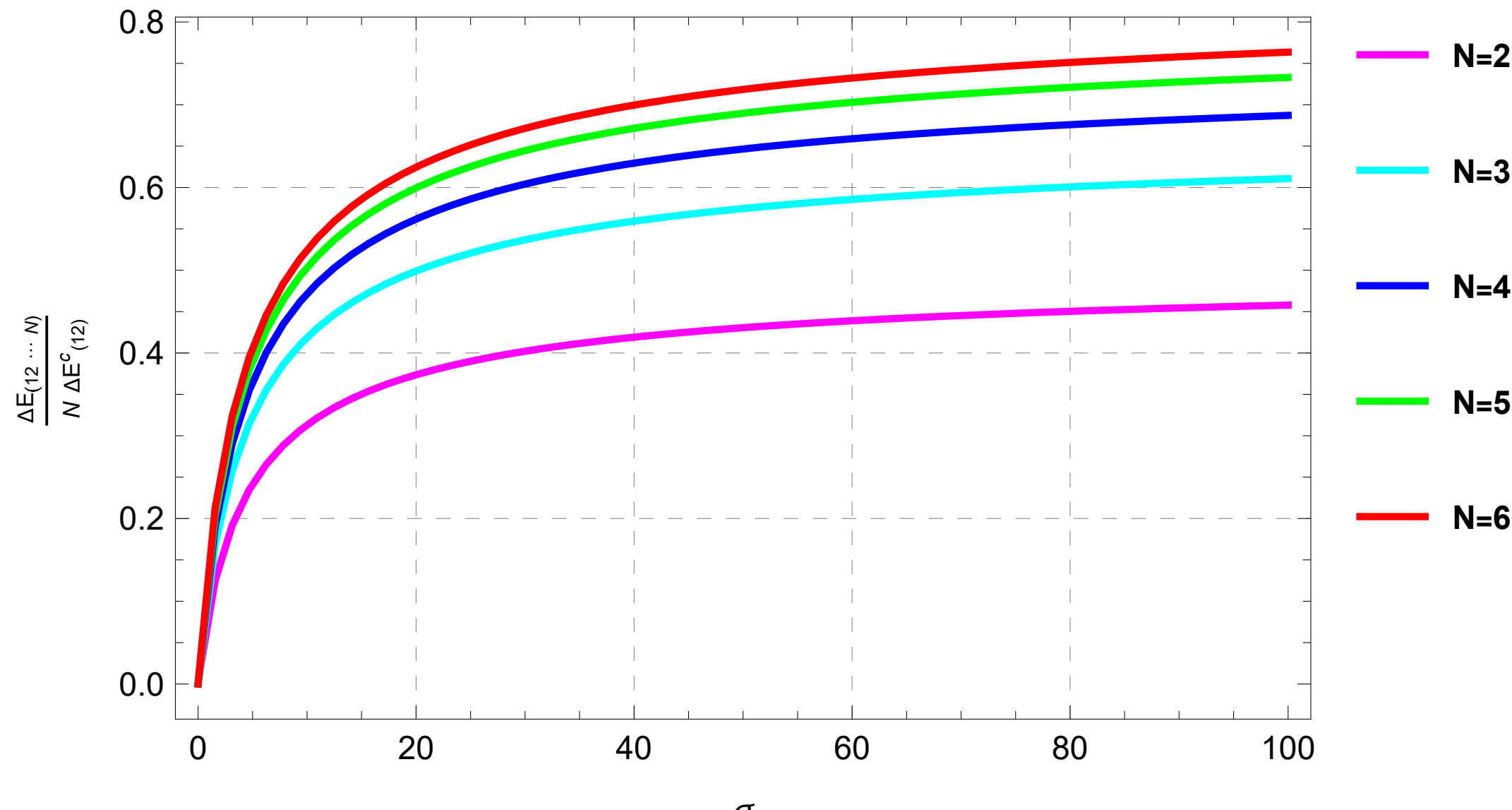

FIG. 2: Casimir energy of $N = 2, 3, 4, 5, 6$ plates scaled by $E^c_{(12)}$ and normalised per unit plate $N$ for varying $\sigma$.

(In the asymptotic limit $\sigma \to \infty$) as also shown in Ref. [35]. Consecutively $\Delta E^c_{(12\cdots N)}/N \to \Delta E^c_{(12)}$ for $N >> 1$.

## IV. CASIMIR ENERGY BETWEEN INFINITELY PERMEABLE AND DIELECTRIC $\delta$-PLATES

In direct contrast to Casimir's result of attractive force [1], Boyer discovered repulsive force from the interaction between a pure dielectric material and a pure magnetic conductor as $\Delta E^b_{(12)} = -7/8\Delta E^c_{(12)}$ [50]. This is evident from the optical coefficients of $\delta$ plate with an infinitely permeable material from Eq. (2) where $r_g^H = -1, r_g^E = 1$ and $t_g^{H,E} = 0$ for $\lambda_g^\perp \to \infty$, whereas optical coefficients of $\delta$ plate with an infinitely dielectric material are $r_e^H = 1, r_e^E = -1$ and $t_e^{H,E} = 0$ for $\lambda_e^\perp \to \infty$. Using this the Casimir energy of $N = 2$ plates from Eq. (1) with $r_1 = r_e$, $r_2 = r_g$, $r_{3,4,5,6} = 0$ and $l_{ij} = a$ in Eq. (7) leads to

$$\frac{\Delta E^b_{(12)}}{A} = \frac{1}{2}\int_{-\infty}^{\infty}\frac{d\zeta}{2\pi}\int\frac{d^2k_\perp}{(2\pi)^2}\left[\ln\left[1+e^{-2\kappa a}\right]+\ln\left[1+e^{-2\kappa a}\right]\right]. \tag{16}$$

Evaluating this gives $\Delta E^b_{(12)}/A = 7/8(\pi^2/720a^3)$ leading to a repulsive force. This force is repulsive since $\mathbf{F_1}\cdot\hat{\mathbf{z}}/A = -7/8(\pi^2/240a^4)$ and $\mathbf{F_2}\cdot\hat{\mathbf{z}}/A = 7/8(\pi^2/240a^4)$ from Eq. (8). While the magnitude of this force decreases with increasing $a$, its direction remains unchanged.

Interestingly, the Casimir energy for $N$ $\delta$ plates with alternating pure electric and magnetic properties (Zaremba boundary conditions with alternating Dirichlet and Neumann boundary conditions in scalar case [51]) also leads to

$$\Delta E^b_{(12\cdots N)} = (N-1)\Delta E^b_{(12)} \tag{17}$$

since transmission coefficients $t_g^{H,E} = 0$ and $t_e^{H,E} = 0$, where next-to-next-to-nearest neighbour, $\cdots$ scattering parameters $\Delta_{ik}$ in Eq. (5) do not contribute.

Also, while Casimir energy of $N = 2$ plates from Eq. (1) with $r_1 = r_e$, $r_2 = r_\sigma$, $r_{3,4,5,6} = 0$ and $l_{ij} = a$ in Eq. (7) leads to

$$\frac{\Delta E_{(12)}}{A} = \frac{1}{2}\int_{-\infty}^{\infty}\frac{d\zeta}{2\pi}\int\frac{d^2k_\perp}{(2\pi)^2}\left[\ln\left[1-\left(\frac{\sigma\kappa}{\sigma\kappa+2\zeta}\right)e^{-2\kappa a}\right]+\ln\left[1-\left(\frac{\sigma\zeta}{\sigma\zeta+2\kappa}\right)e^{-2\kappa a}\right]\right], \tag{18}$$

and evaluating with $\sigma = \pi\alpha$ gives an attractive force of $\Delta E_{(12)} = 0.027\Delta E^c_{(12)}$ between a Graphene plate and perfect dielectric material. Whereas, Casimir energy from Eq. (1) with $r_1 = r_g$, $r_2 = r_\sigma$, $r_{3,4,5,6} = 0$ and constant $l_{ij} = a$ in Eq. (7) leads to

$$\frac{\Delta E_{(12)}}{A} = \frac{1}{2}\int_{-\infty}^{\infty}\frac{d\zeta}{2\pi}\int\frac{d^2k_\perp}{(2\pi)^2}\left[\ln\left[1+\left(\frac{\sigma\kappa}{\sigma\kappa+2\zeta}\right)e^{-2\kappa a}\right]+\ln\left[1+\left(\frac{\sigma\zeta}{\sigma\zeta+2\kappa}\right)e^{-2\kappa a}\right]\right], \tag{19}$$

and evaluating it gives an repulsive force of $\Delta E_{(12)} = -0.026\Delta E^c_{(12)}$ between a Graphene plate and perfect magnetic permeable material. The energy leads to repulsive force whenever $r_1^{H(E)} r_2^{H(E)} < 0$.

Further, we study the effect of an infinitely permeable plate within a stack and outside a stack of finitely conducting plates in Casimir energy of $N = 3, 4, 5, 6$ from Eq. (1). Casimir energy of $N = 3$ plates with $r_1 = r_g$, $r_{2,3} = r_\sigma$, $r_{4,5,6} = 0$ and $r_2 = r_g$, $r_{1,3} = r_\sigma$, $r_{4,5,6} = 0$ with constant $l_{ij} = a$ in Eq. (7) for varying conductivity $\sigma$ is displayed in Fig. 3. We observe that for small values of $\sigma$ the perfectly permeable plate adjacent to two consecutive constant conductivity plates causes repulsion force and as $\sigma$ increases the force turns attractive. Meanwhile, the force is always repulsive when the perfectly permeable plate is between constant conductivity plates, as it physically disconnects the two spaces. From Eq. (9), it is evident that while the magnitude of the force on the first and last plates in the stack decreases with increasing separation $a$, its direction remains unchanged.

In the case of Casimir energy of $N = 4$ plates with a perfectly permeable plate adjacent to three constant conductivity plates ($r_1 = r_g$, $r_{2,3,4} = r_\sigma$), the force is always attractive. With a perfectly permeable plate in between three constant conductivity plates ($r_2 = r_g$, $r_{1,3,4} = r_\sigma$), the force is always repulsive as shown in Fig. 4. Similarly, the Casimir energy of $N = 5, 6$ plates are shown in Figs. 5, 6, where the perfectly permeable plate inside the stack causes a repulsive force only for smaller values of $\sigma$.

In the asymptotic limit for perfect metals ($\sigma \to \infty$), the values of Casimir energy can be easily evaluated from Eqs. (15) and (17). For instance, the Casimir energy for $N = 4$ when $\sigma \to \infty$ is $\Delta E_{(1234)} = (\Delta E^b_{(12)} + \Delta E^c_{(23)} + \Delta E^c_{(34)})/4\Delta E^c_{(12)} = 0.28125$ for $r_1 = r_g$, $r_{2,3,4} = r_\sigma$ and $\Delta E_{(1234)} = (\Delta E^b_{(12)} + \Delta E^b_{(23)} + \Delta E^c_{(34)})/4\Delta E^c_{(12)} = -0.1875$ for $r_2 = r_g$, $r_{1,3,4} = r_\sigma$ as can be seen in Fig. 4. The corresponding force for $N = 4$ when $\sigma \to \infty$ is $\mathbf{F_1}\cdot\hat{\mathbf{z}}/A = (1.125)\pi^2/240a^4$, $\mathbf{F_4}\cdot\hat{\mathbf{z}}/A = -(1.125)\pi^2/240a^4$ for $r_1 = r_g$, $r_{2,3,4} = r_\sigma$ and $\mathbf{F_1}\cdot\hat{\mathbf{z}}/A = -(0.75)\pi^2/240a^4$, $\mathbf{F_4}\cdot\hat{\mathbf{z}}/A = (0.75)\pi^2/240a^4$ for $r_2 = r_g$, $r_{1,3,4} = r_\sigma$ from Eq. (8).

### A. Neural network analysis of the Casimir energy

We implemented the MATLAB fitnet feedforward neural network [52] specifically designed for regression task such as to identify a relationship between normalized many-body energy $\Delta E_{(12\cdots N)}/(N\Delta E^c_{(12)})$ and the scaled pairwise energies $\{\Delta E_{(12)}/\Delta E^c_{(12)}, \Delta E_{(23)}/\Delta E^c_{(12)}, \cdots, \Delta E_{(N-1,N)}/\Delta E^c_{(12)}\}$. The neural network architecture is visualized in Fig. 7.

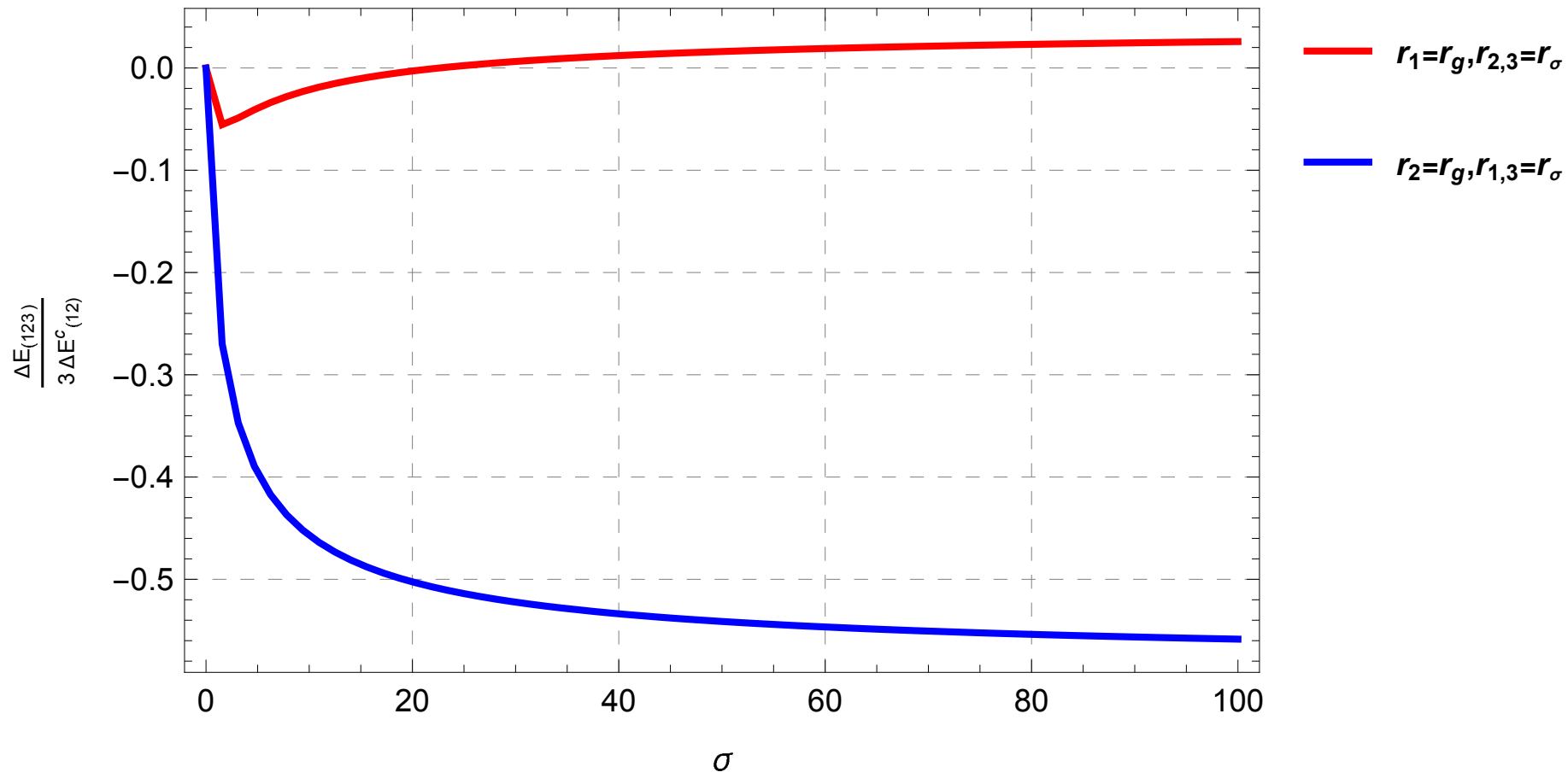

FIG. 3: Casimir energy of $N = 3$ plates with an infinitely permeable plate and for varying $\sigma$ of other plates.

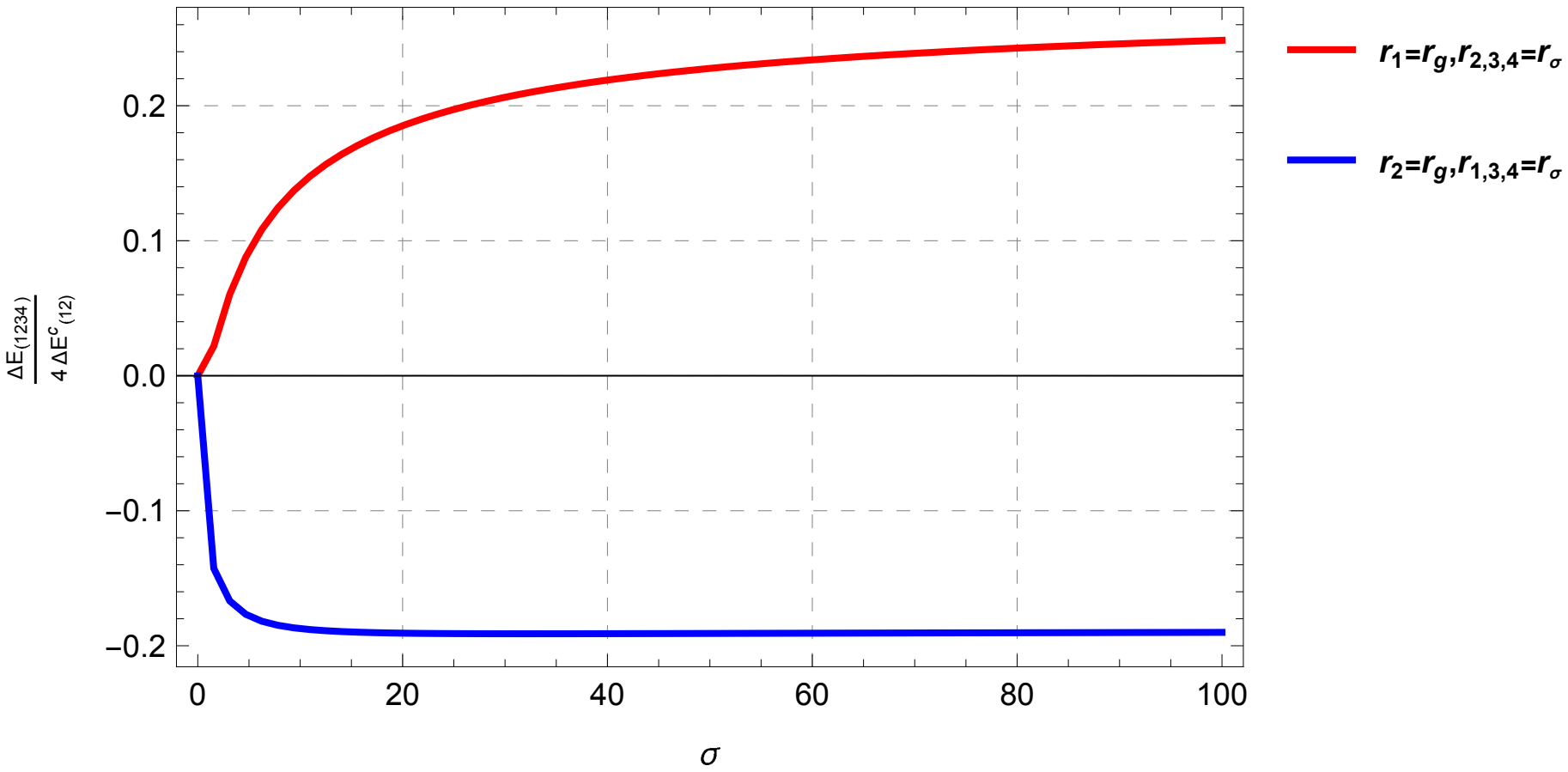

FIG. 4: Casimir energy of $N = 4$ plates with an infinitely permeable plate and for varying $\sigma$ of other plates.

Key characteristics of the network we implement include an input layer which accepts two numerical datasets derived from $\Delta E(r_\sigma, r_\sigma)$ (Energy of configuration with $r_1 = r_\sigma$, $r_2 = r_\sigma$, $r_{3,4,5,6} = 0$ and $l_{ij} = a$ in Eq. (7) scaled by $\Delta E^c_{(12)}$) and $\Delta E(r_g, r_\sigma)$ (Energy of configuration with $r_1 = r_g$, $r_2 = r_\sigma$, $r_{3,4,5,6} = 0$ and $l_{ij} = a$ in Eq. (7) scaled by $\Delta E^c_{(12)}$). The intermediate hidden layer contains ten neurons with a Hyperbolic Tangent Sigmoid activation function. Tansig function is defined as

$$f(x) = \frac{2}{1 + e^{-2x}} - 1, \tag{20}$$

which captures the input values in the range $[-1, 1]$, making it suitable for both positive and negative normalized input data, as in our case. Furthermore, the output layer consists of a single neuron with a purelin (linear) activation function, suitable for regression. Purelin function such as $f(x) = x$, passes the weighted sum of pairwise energy inputs from the hidden layer to learn the normalized and scaled many-body energy output $\Delta E_{(12\cdots N)}/(N\Delta E^c_{(12)})$. Further, the neural network is trained using the Levenberg-Marquardt backpropagation algorithm. According to the algorithm, the input data propagates through the network, and the predicted output ($\hat{E}$) is calculated. The performance of the network is evaluated using the mean squared error (mse):

$$\text{mse} = \frac{1}{n}\sum_{i=1}^{n}(E_i - \hat{E}_i)^2,$$

where $E_i$ is the true energy output and $\hat{E}_i$ is the predicted energy output with $n$ indicating the discrete data points for varying $\sigma$. In each epoch (one instance of learning from the inputs to predict the output), the gradients of the mse with respect to network weights are computed, and the weights $w$ of each neuron are updated as:

$$w - \eta \frac{\partial \text{mse}}{\partial w} \rightarrow w,$$

where $\eta$ is the learning rate of the network. The weights are updated to minimize the mse iteratively. The network learns a set of weights $w$ that connect inputs to the hidden layer. These weights indicate the relative importance of each input feature. The weight's magnitude reflects the corresponding input's contribution to the network's predicted output.

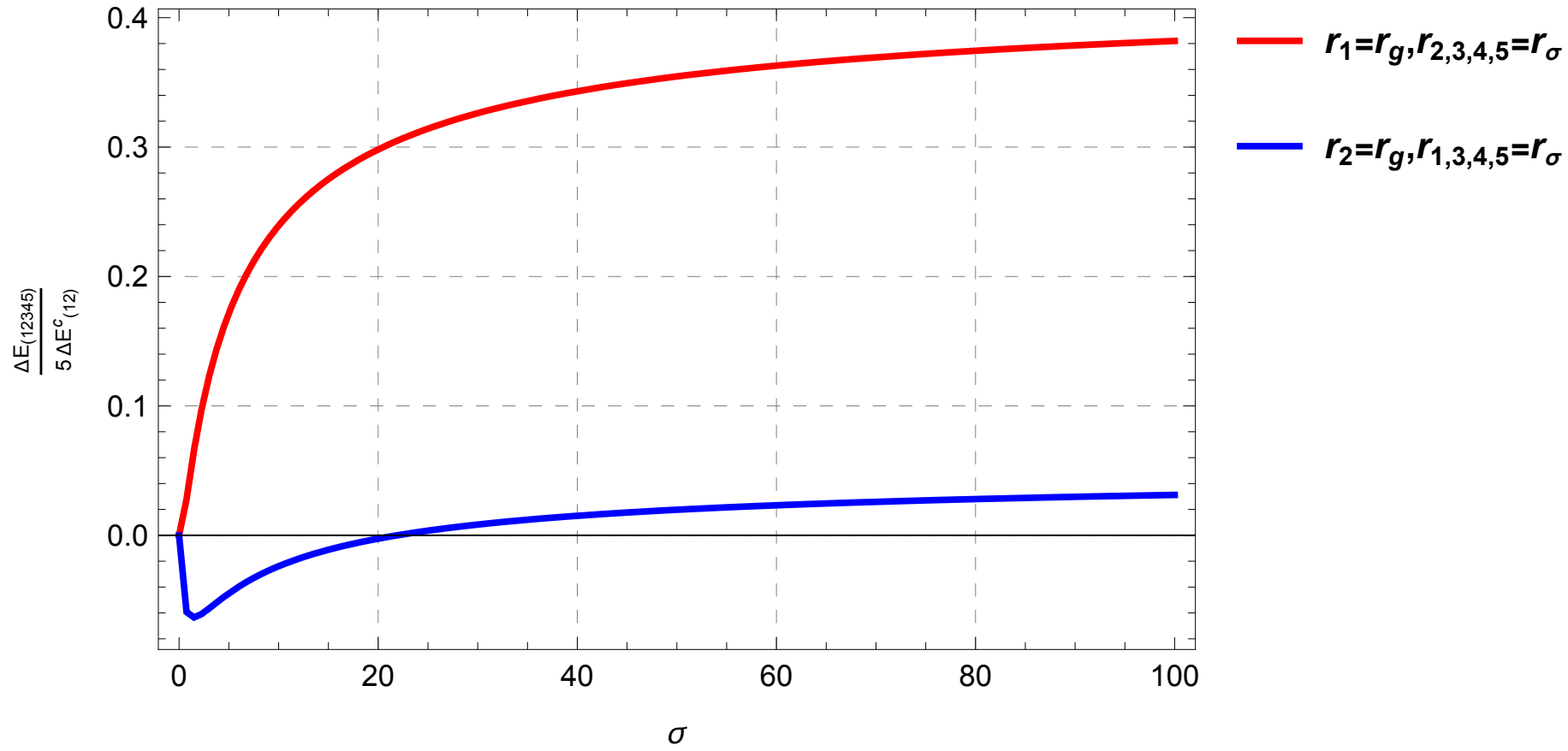

FIG. 5: Casimir energy of $N = 5$ plates with an infinitely permeable plate and for varying $\sigma$ of other plates.

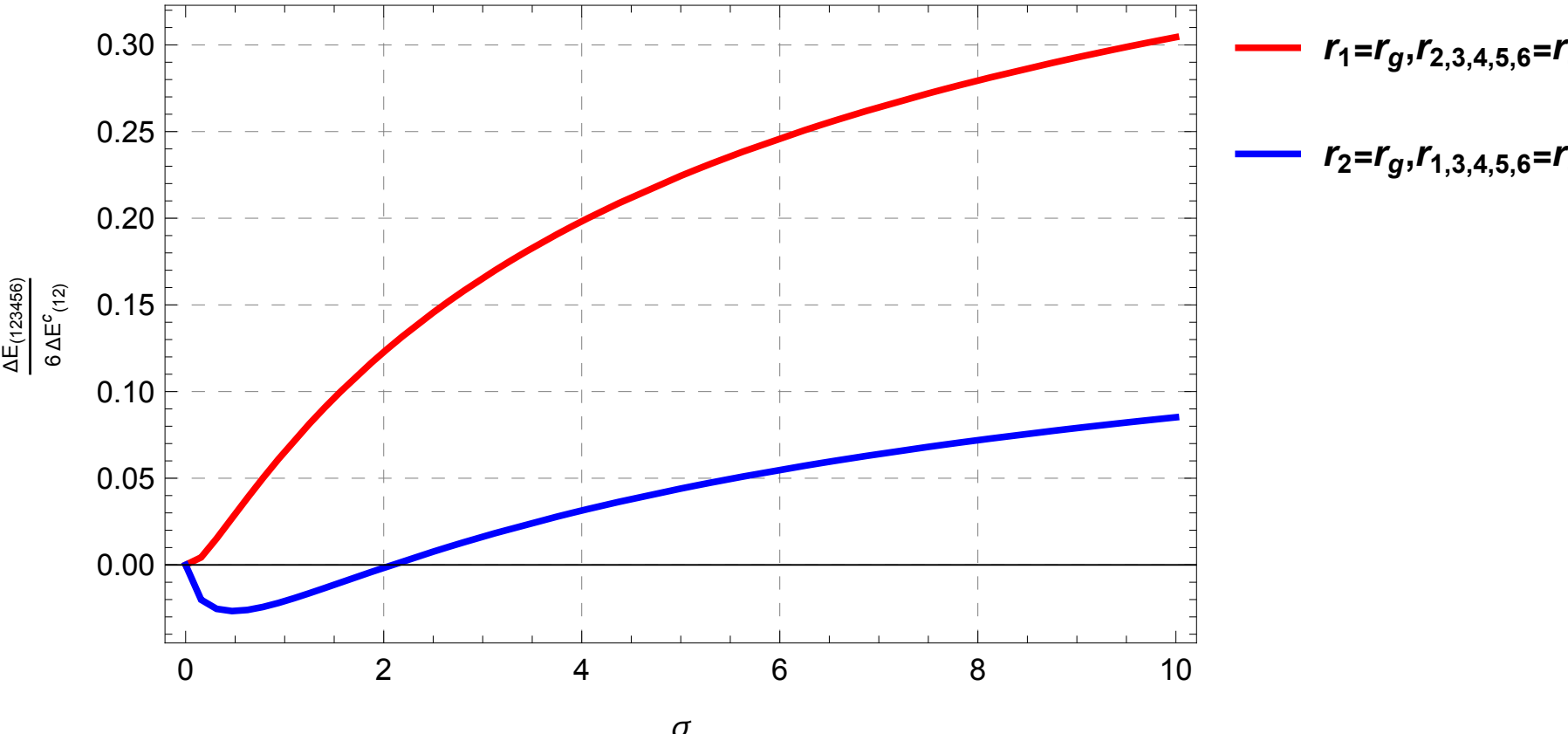

FIG. 6: Casimir energy of $N = 6$ plates with an infinitely permeable plate and for varying $\sigma$ of other plates.

We initially illustrate the behavior of weights in the neurons for a particular instance of output considered as $\Delta E_{(123)}/(3\Delta E^c_{(12)})$ (studied in Fig. 3) with $r_1 = r_g$, $r_{2,3} = r_\sigma$, $r_{4,5,6} = 0$ for conductivity varying from $0 < \sigma < 20$ (region of repulsive force) and $0 < \sigma < 100$ ($20 < \sigma < 100$ is region of attractive force) in Fig. 8. We show the visualization of updated weights of the pairwise energy inputs (Figs. 8(a),(b)) for the lowest mse from 100 different runs (Fig. 8(c),(d)) when the neural network is made to learn based on the many-body energy output. This visualization provides a detailed understanding of input features $\Delta E(r_g, r_\sigma)$ and $\Delta E(r_\sigma, r_\sigma)$ in the neural network's decision-making process. The variation in positive and negative weights across hidden nodes suggests that the network has learned different patterns of positive and negative influence for each input feature. For the run with the lowest mse, we observe that the contribution of $\Delta E(r_\sigma, r_\sigma)$ to the hidden layer nodes has consistently higher positive weight magnitudes across most nodes and has a more substantial influence on the network's output for conductivity varying from $0 < \sigma < 100$ indicating the region is mostly of attractive force. Whereas, $\Delta E(r_g, r_\sigma)$ has a prominent influence on nodes with positive weights of higher magnitudes than the weight of $\Delta E(r_\sigma, r_\sigma)$ for conductivity varying from $0 < \sigma < 20$, indicating the region is mostly of repulsive force.

Similarly we study the weights learned from the neural networks for numerical data of Casimir energy from $\Delta E_{(1234)}/(4\Delta E^c_{(12)})$, $\Delta E_{(12345)}/(5\Delta E^c_{(12)})$ and $\Delta E_{(123456)}/(6\Delta E^c_{(12)})$ studied in Fig.s 4, 5 and 6, respectively. By summing the values of input weights over all the hidden neurons, we can compute an aggregated importance score for each input $\Delta E(r_\sigma, r_\sigma)$ and $\Delta E(r_g, r_\sigma)$ as displayed in Fig.s 9 and 10. We observe that the neural network identifies the region of attractive and repulsive forces from the relative average weights. For instance, the configuration in considered in Fig. 6, $\Delta E_{(123456)}/(6\Delta E^c_{(12)})$ with $r_2 = r_g$, $r_{1,3,4,5} = r_\sigma$ has region of repulsive force when $0 < \sigma < 2$, which can be interpreted by the average weight of $\Delta E(r_g, r_\sigma)$ slightly higher than average weight of $\Delta E(r_\sigma, r_\sigma)$ in Fig. 10(c).

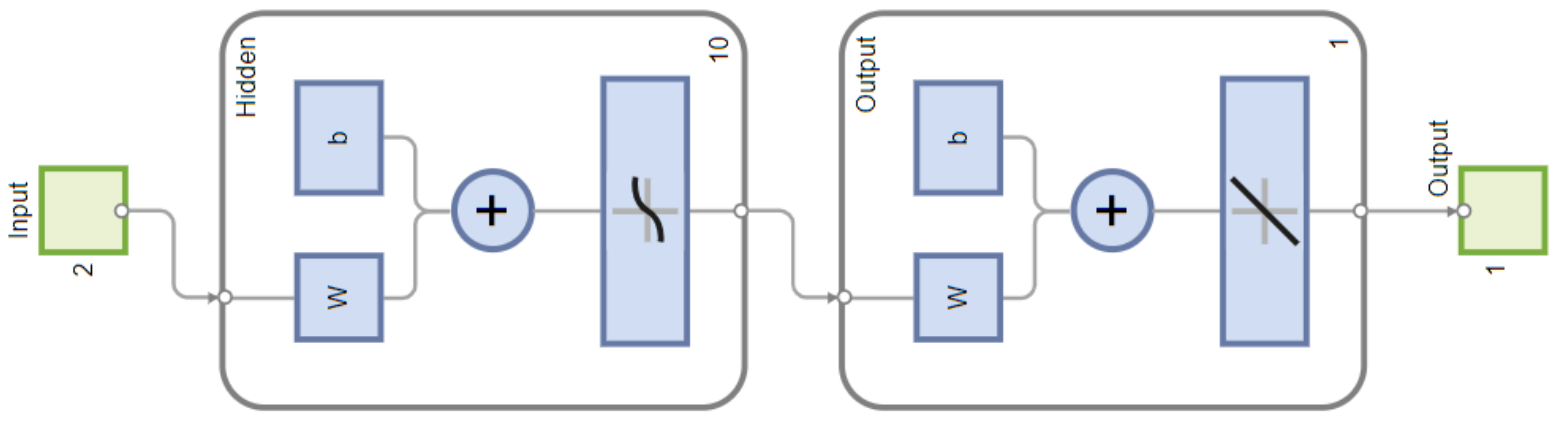

FIG. 7: MATLAB neural network architecture with inputs as numerical data from $\Delta E(r_\sigma, r_\sigma)$ and $\Delta E(r_g, r_\sigma)$ for varying $\sigma$ with intermediate hidden layer (10 Neurons) and output is the numerical data from $\Delta E_{(12\cdots N)}/(N\Delta E^c_{(12)})$ for corresponding $\sigma$.

## V. CONCLUSIONS

The closed-form expression of Casimir energy and corresponding force for $N = 6$ $\delta-$function plates was derived depending on multiple scattering parameter $\Delta$; the result for $N$ plates may be easily generalized as described. Casimir energy of $N = 2, \cdots, 6$ plates with Graphene-related constant conductivity was computed and numerically studied, comparing with existing studies [16, 23]. Because of Graphene's transparency and finite optical conductivity, we found that the Casimir energy between Graphene plates obeys the same distance dependence as the energy between two perfect electric conductors but with significantly smaller magnitudes.

We also studied the interaction of infinitely permeable material and $\delta$ plates with constant conductivity, which results in repulsive Casimir forces. We observed the change in repulsive to attractive forces based on the configuration and position of an infinitely permeable $\delta$ plate in the Casimir energy of $N = 3, \cdots, 6$ plates. To complement these numerical results, we used a feedforward neural network in MATLAB to analyze the many-body energy across configurations. The neural network captured the influence of pairwise negative and positive energies, identifying regions of attractive and repulsive forces in the many-body energy. The neural network's weights reflected the transitions, demonstrating its robustness and reliability in modeling many-body interactions.

Further, experimentally relevant parameters like binding energies [22, 23] can perhaps be studied using the Drude-Lorentz model of conductivity for Graphene [53]. Also, accurate characterization of Graphene using its dynamical conductivity is essential for modeling many optoelectronic devices, such as tunable Graphene plasmonic structures in the mid-infrared to terahertz regime [54–56]. In this context, Graphene's optical response has been extensively studied using its spatiotemporal conductivity tensor derived via methods like the polarization tensor [57, 58], semi-classical Boltzmann transport [59], and the Kubo formalism [60]. Complementing these continuum models, atomistic first-principles methods (especially density functional theory) were also recently applied to Graphene and related two-dimensional materials [61]. These ab initio approaches can compute binding energies and van der Waals dispersion forces that are difficult to capture with simpler models. For instance, Ref. [62] compared density functional theory and macroscopic quantum electrodynamics calculations for van der Waals binding in Fullerene (a carbon-based nanostructure closely related to Graphene) and found that the two methods produce consistent results. Similarly, a novel second-harmonic Hall response was demonstrated recently in multilayered 2D materials via density functional theory – an emergent phenomenon not predicted by typical conductivity models [63].

The attractive force between two plates with dielectric permittivity $\epsilon_1 \to \infty$ and $\epsilon_2 \to \infty$ was determined by Casimir [1]. It was further generalized for two dielectric slabs with $\epsilon_3$ in the space filled between them and for arbitrary values of $\epsilon_1$ and $\epsilon_2$ [64]. When similar bodies $\epsilon_1 = \epsilon_2 \equiv \epsilon$ are considered, this result yields an attractive force, independent of $\epsilon_3$. However, it was confirmed experimentally that the force can become repulsive when a dissimilarity occurs, such as $\epsilon_1 > \epsilon_3 > \epsilon_2$ in the dielectric characteristics [65, 66]. Repulsion can be realized in

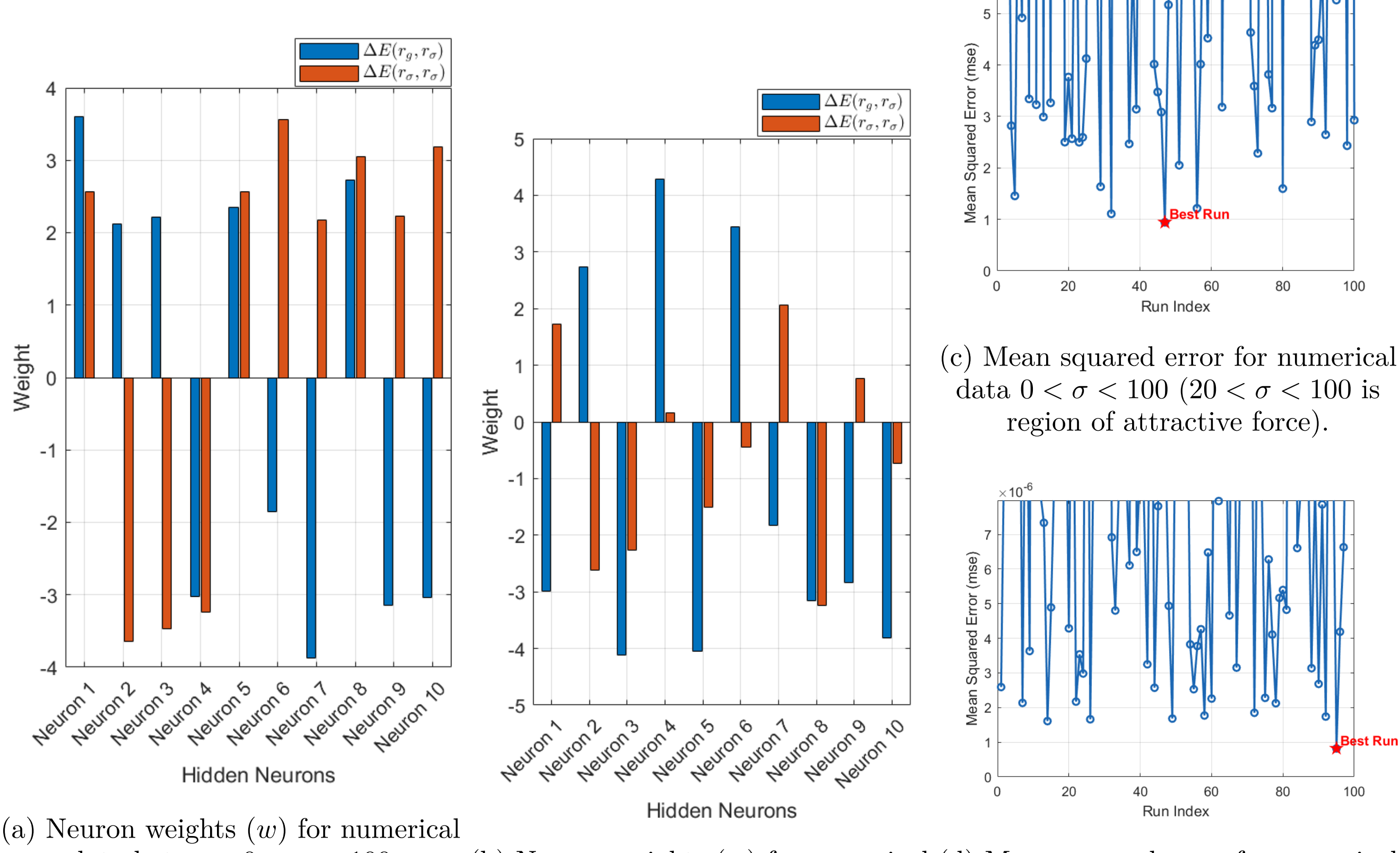

(a) Neuron weights ($w$) for numerical data between $0 < \sigma < 100$ ($20 < \sigma < 100$ is region of attractive force).

(b) Neuron weights ($w$) for numerical data between $0 < \sigma < 20$ (region of repulsive force).

(c) Mean squared error for numerical data $0 < \sigma < 100$ ($20 < \sigma < 100$ is region of attractive force).

(d) Mean squared error for numerical data $0 < \sigma < 20$ (region of repulsive force).

FIG. 8: (a) and (b) Visualization of neuron weights for the inputs $\Delta E(r_\sigma, r_\sigma)$ and $\Delta E(r_g, r_\sigma)$ with output as $\Delta E_{(123)}/(3\Delta E^c_{(12)})$ for data with varying conductivity $\sigma$. The x-axis represents the hidden layer neurons, and the y-axis represents the weight values. Positive weights indicate a positive influence, while negative weights suggest a negative influence on the hidden nodes. (c) and (d) The corresponding mean squared error (mse) over 100 different runs of neural networks. After learning, the lowest mses are in the range of $10^{-6}$.

other contexts of conventional electrodynamics, such as employing magnetically permeable materials [67] comparable to Boyer [50], and from shifting geometries [68], beyond the Casimir-Lifshitz results only using dielectrics. Recent experiments also witnessed repulsion using large external magnetic fields [69]. Boyer repulsion is interesting [67], but since naturally occurring materials do not exhibit strong magnetic responses, it has been regarded as non-physical and challenging [70, 71]. However, recent advancements in nanofabrication have produced metamaterials with magnetic responses [72], which may be relevant to the physical realization of Boyer's repulsion [73–76]. Magnetic-response metamaterial substrates have also been studied in conjunction with Graphene to explore repulsive Casimir forces [77]. Recently, magnetic van der Waals materials [78, 79] have also been investigated to realize Boyer repulsion [80]. Also, understanding the fundamental notion of electromagnetic force density turning negative (attractive force) to positive (repulsive force) is nontrivial [48, 81], which needs further investigation.

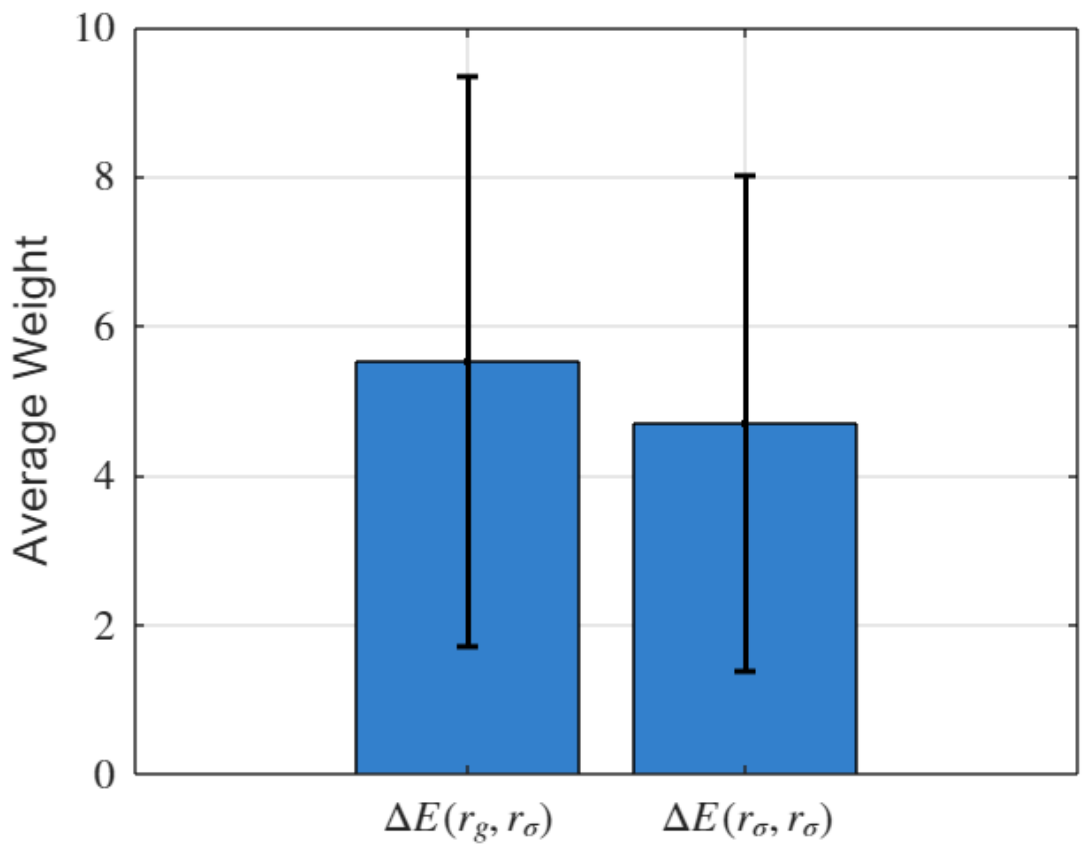

(a) Weights for output as $\Delta E_{(1234)}/(4\Delta E^c_{(12)})$ with $r_2 = r_g$, $r_{1,3,4} = r_\sigma$ with $0 < \sigma < 100$.

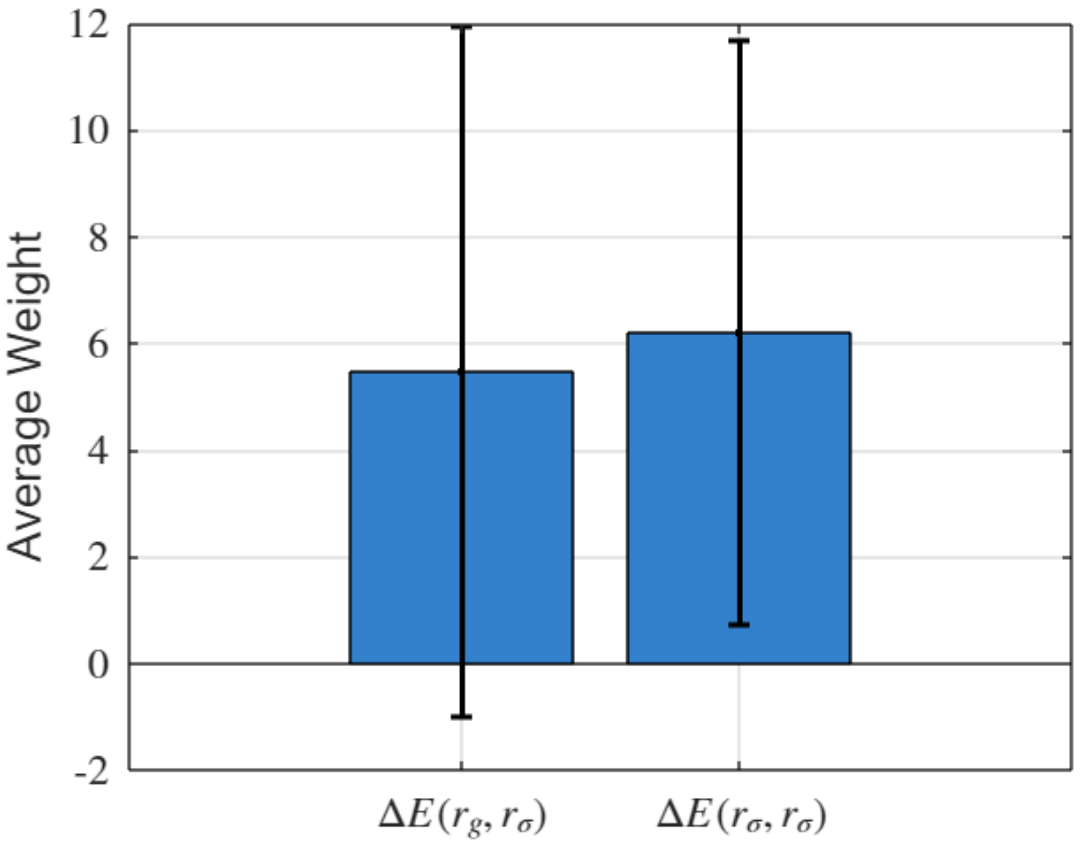

(b) Weights for output as $\Delta E_{(1234)}/(4\Delta E^c_{(12)})$ with $r_1 = r_g$, $r_{2,3,4} = r_\sigma$ with $0 < \sigma < 100$.

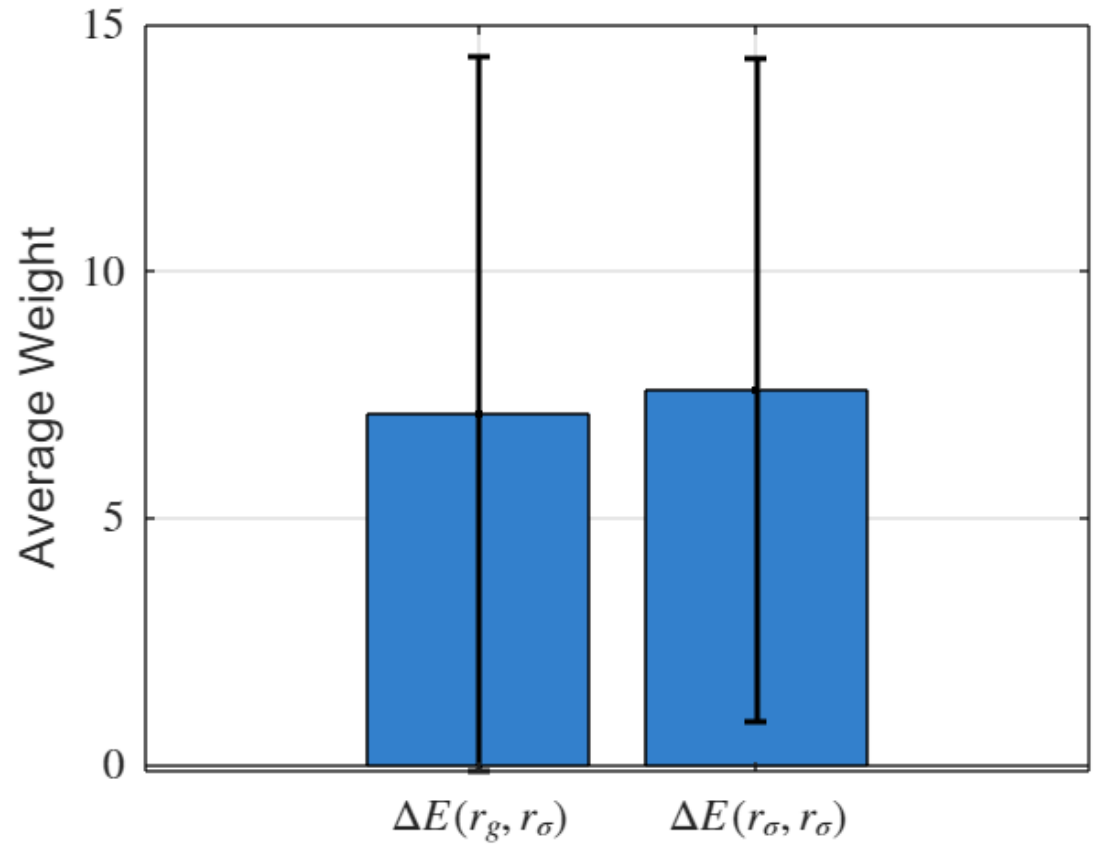

(c) Weights for output as $\Delta E_{(12345)}/(5\Delta E^c_{(12)})$ with $r_2 = r_g$, $r_{1,3,4,5} = r_\sigma$ with $0 < \sigma < 100$.

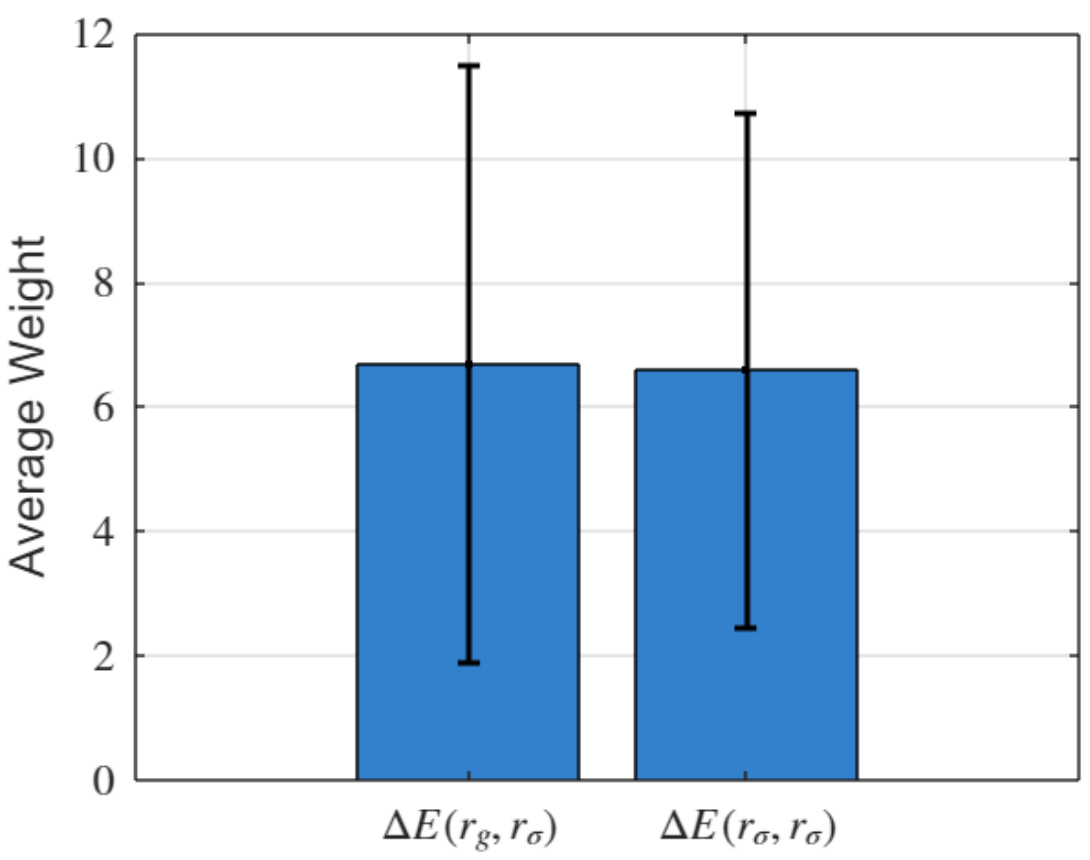

(d) Weights for output as $\Delta E_{(12345)}/(5\Delta E^c_{(12)})$ with $r_2 = r_g$, $r_{1,3,4,5} = r_\sigma$ with $0 < \sigma < 20$.

FIG. 9: *Averaged weights of inputs $\Delta E(r_g, r_\sigma)$ and $\Delta E(r_\sigma, r_\sigma)$ with bars showing variations over neurons.*

## VI. APPENDIX

The Casimir energy of $N = 3$ plates from Eq. (1) is

$$\frac{\Delta E_{(123)}}{A} = \frac{1}{2}\int_{-\infty}^{\infty}\frac{d\zeta}{2\pi}\int\frac{d^2k_\perp}{(2\pi)^2}\Bigg[\ln\Big[(1-r_\sigma^H r_\sigma^H e^{-2\kappa a})(1-r_\sigma^H r_\sigma^H e^{-2\kappa a}) - (r_\sigma^H (t_\sigma^H)^2 r_\sigma^H)e^{-4\kappa a}\Big] + \ln\Big[(1-r_\sigma^E r_\sigma^E e^{-2\kappa a})(1-r_\sigma^E r_\sigma^E e^{-2\kappa a}) - (r_\sigma^E (t_\sigma^E)^2 r_\sigma^E)e^{-4\kappa a}\Big]\Bigg] \tag{21}$$

from considering $r_{1,2,3} = r_\sigma$, $t_{1,2,3} = t_\sigma$, $r_{4,5,6} = 0$ and constant $l_{ij} = a$ in Eq. (7). Introducing the optical coefficients of $\sigma$ conductivity $\delta$-plate $r_\sigma, t_\sigma$ from Eq. (11) and expanding inside the logarithm, we obtain

$$\frac{\Delta E_{(123)}}{A} = \frac{1}{2}\int_{-\infty}^{\infty}\frac{d\zeta}{2\pi}\int\frac{d^2k_\perp}{(2\pi)^2}\Bigg[\ln\left(1-2\left(\frac{\sigma\kappa}{\sigma\kappa+2\zeta}\right)^2 e^{-2\kappa a} + \left(\left(\frac{\sigma\kappa}{\sigma\kappa+2\zeta}\right)^4 - \left(\frac{\sigma\kappa}{\sigma\kappa+2\zeta}\right)^2\left(1-\frac{\sigma\kappa}{\sigma\kappa+2\zeta}\right)^2\right)e^{-4\kappa a}\right) + \ln\left(1-2\left(-\frac{\sigma\zeta}{\sigma\zeta+2\kappa}\right)^2 e^{-2\kappa a} + \left(\left(-\frac{\sigma\zeta}{\sigma\zeta+2\kappa}\right)^4 - \left(-\frac{\sigma\zeta}{\sigma\zeta+2\kappa}\right)^2\left(1-\frac{\sigma\zeta}{\sigma\zeta+2\kappa}\right)^2\right)e^{-4\kappa a}\right)\Bigg]. \tag{22}$$

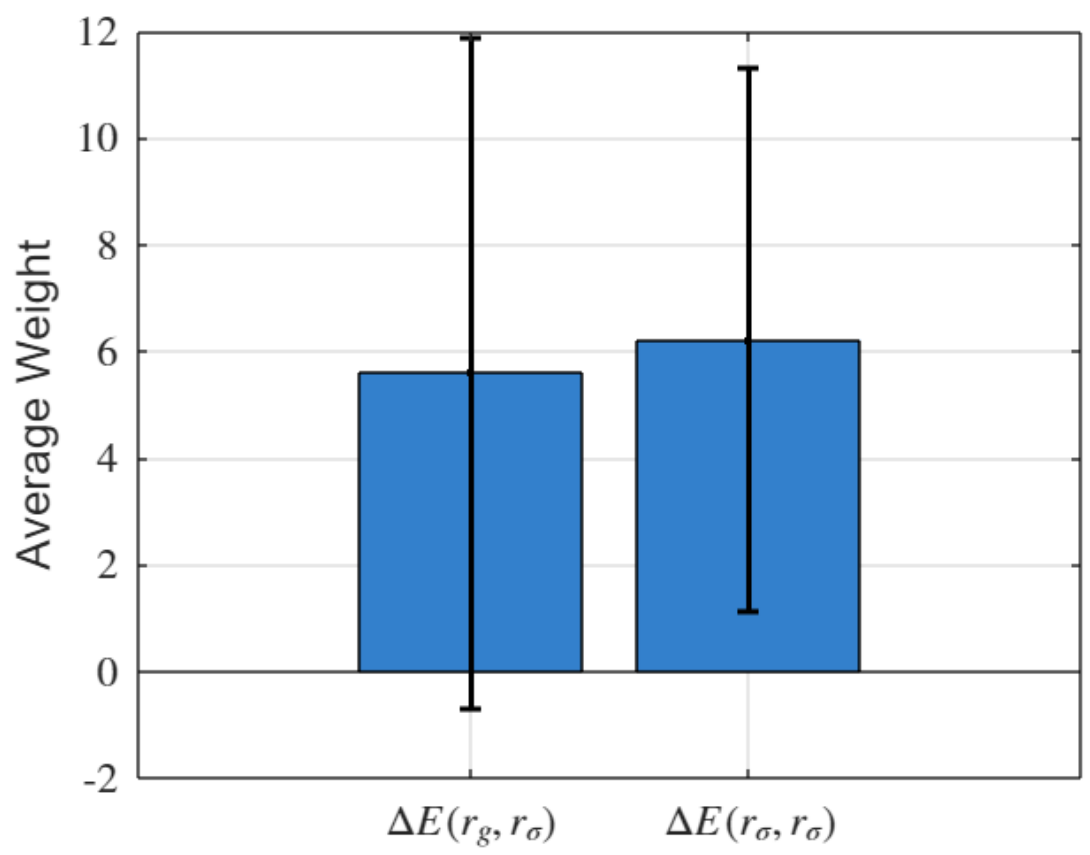

(a)Weights for output as $\Delta E_{(12345)}/(5\Delta E^c_{(12)})$ with $r_1 = r_g$, $r_{2,3,4,5} = r_\sigma$ with $0 < \sigma < 100$.

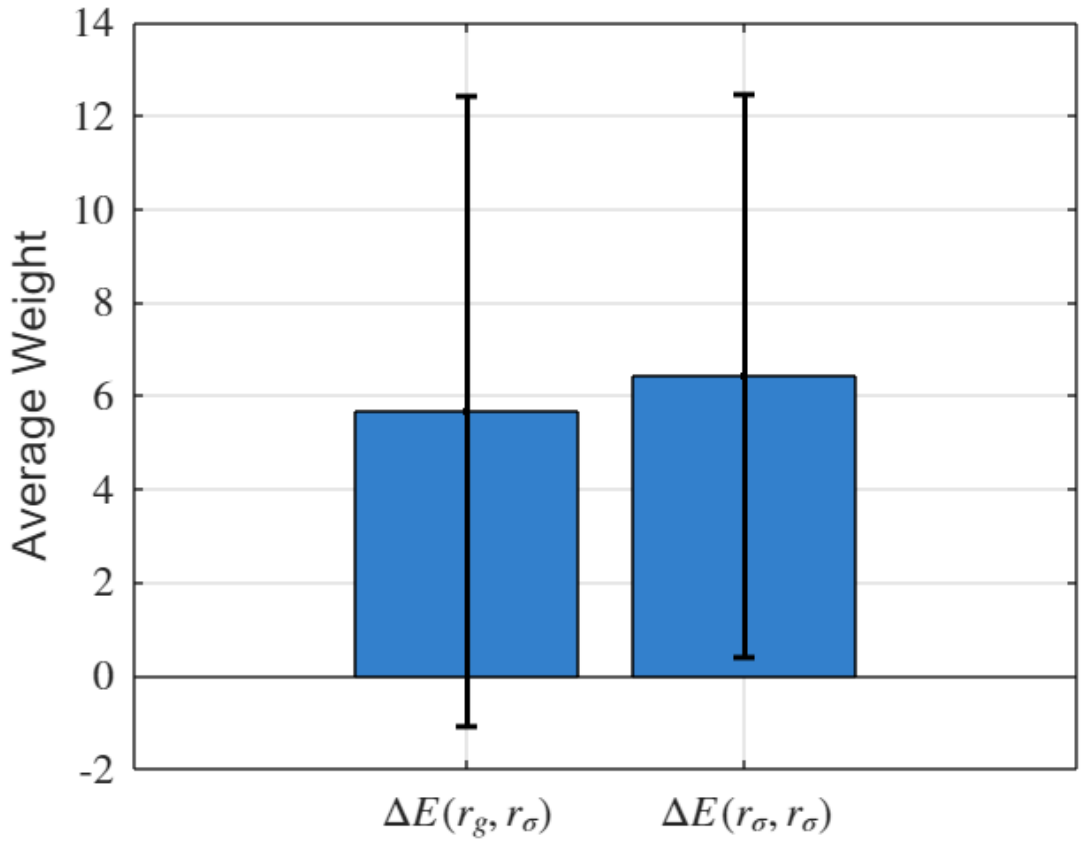

(b) Weights for output as $\Delta E_{(123456)}/(6\Delta E^c_{(12)})$ with $r_2 = r_g$, $r_{1,3,4,5,6} = r_\sigma$ with $0 < \sigma < 10$.

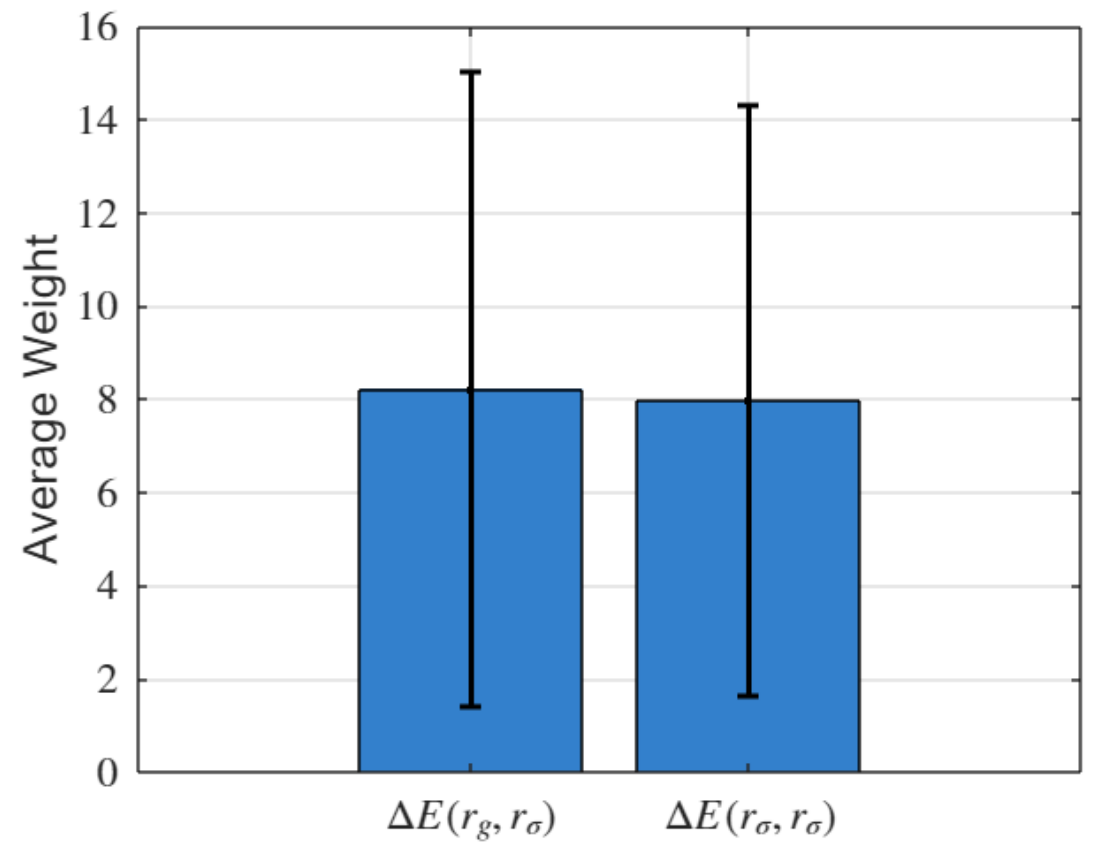

(c) Weights for output as $\Delta E_{(123456)}/(6\Delta E^c_{(12)})$ with $r_2 = r_g$, $r_{1,3,4,5,6} = r_\sigma$ with $0 < \sigma < 2$.

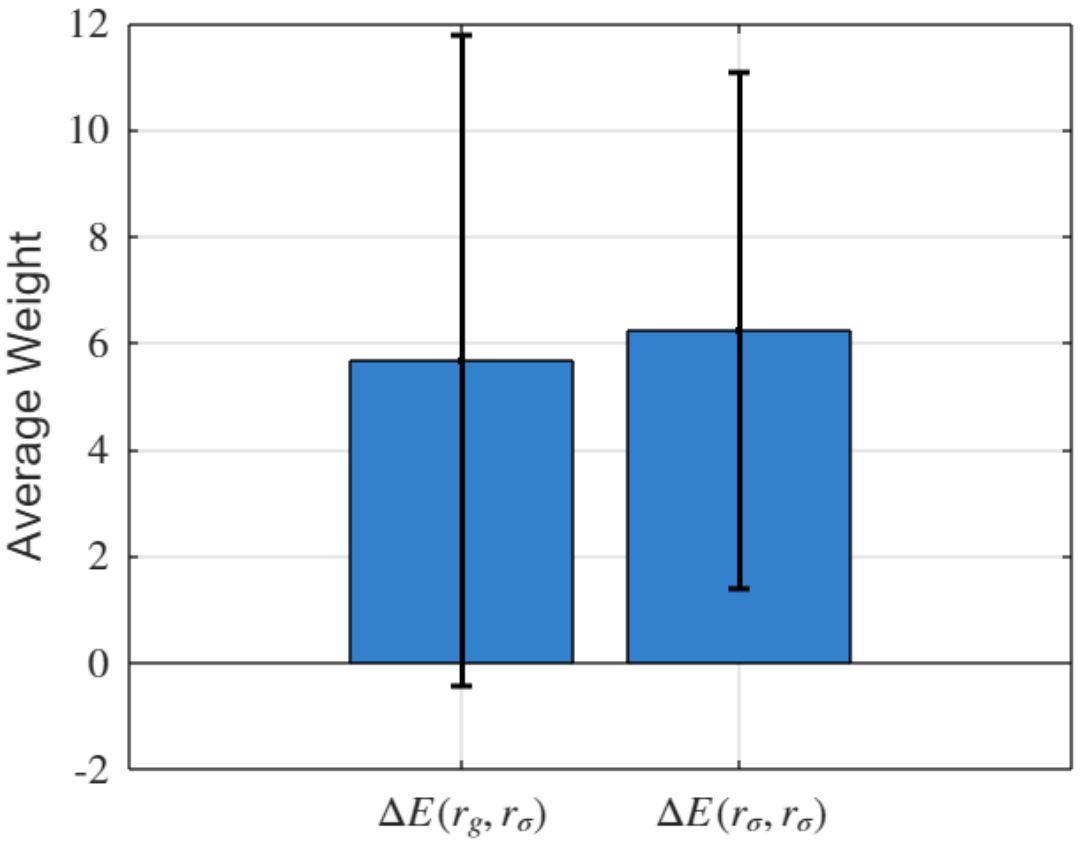

(d) Weights for output as $\Delta E_{(123456)}/(6\Delta E^c_{(12)})$ with $r_1 = r_g$, $r_{2,3,4,5,6} = r_\sigma$ with $0 < \sigma < 10$.

FIG. 10: *Averaged weights of inputs $\Delta E(r_g, r_\sigma)$ and $\Delta E(r_\sigma, r_\sigma)$ with bars showing variations over neurons.*

Scaling with $\Delta E^c_{(12)}$ and introducing the spherical polar coordinates as defined in Sec.III gives

$$\frac{\Delta E_{(123)}}{\Delta E^c_{(12)}} = \frac{1}{2}\int_{-\infty}^{\infty}\frac{d\zeta}{2\pi}\int\frac{d^2k_\perp}{(2\pi)^2}\left[\ln\left[1-2\left(\frac{\sigma}{\sigma+2t}\right)^2 e^{-2s}+\left(\left(\frac{\sigma}{\sigma+2t}\right)^4-\left(\frac{\sigma}{\sigma+2t}\right)^2\left(1-\frac{\sigma}{\sigma+2t}\right)^2\right)e^{-4s}\right]\right.$$
$$\left.+\ln\left[1-2\left(-\frac{\sigma}{\sigma+\frac{2}{t}}\right)^2 e^{-2s}+\left(\left(-\frac{\sigma}{\sigma+\frac{2}{t}}\right)^4-\left(-\frac{\sigma}{\sigma+\frac{2}{t}}\right)^2\left(1-\frac{\sigma}{\sigma+\frac{2}{t}}\right)^2\right)e^{-4s}\right]\right]. \quad (23)$$

Further, evaluation of $s$ integral gives

$$\frac{\Delta E_{(123)}}{\Delta E^c_{(12)}} = \frac{45}{\pi^4}\int_0^1 dt\left[\sum_{i=1}^{2}\mathrm{Li}_4\left(\frac{1}{\alpha_i}\right)+\sum_{i=1}^{2}\mathrm{Li}_4\left(\frac{1}{\beta_i}\right)\right], \quad (24)$$

where $\alpha_1, \alpha_2$ are the roots of the polynomial equation $1+a_1x+b_1x^2 = 0$, and $\beta_1, \beta_2$ be the roots of $1+a_2x+b_2x^2 = 0$ with

$$a_1 = -2\left(\frac{\sigma}{\sigma+2t}\right)^2,\ a_2 = -2\left(-\frac{\sigma}{\sigma+\frac{2}{t}}\right)^2, \quad (25)$$

are coefficients of $e^{-2s}$ and

$$b_1 = \left(\frac{\sigma}{\sigma+2t}\right)^4 - \left(\frac{\sigma}{\sigma+2t}\right)^2 \left(1-\frac{\sigma}{\sigma+2t}\right)^2,\ b_2 = \left(-\frac{\sigma}{\sigma+\frac{2}{t}}\right)^4 - \left(-\frac{\sigma}{\sigma+\frac{2}{t}}\right)^2 \left(1-\frac{\sigma}{\sigma+\frac{2}{t}}\right)^2 \tag{26}$$

are coefficients of $e^{-4s}$ in Eq. (23). Similarly, the Casimir energy of $N=4$ plates from Eq. (1) is

$$\begin{aligned}\frac{\Delta E_{(1234)}}{A} = \frac{1}{2}\int_{-\infty}^{\infty}\frac{d\zeta}{2\pi}\int\frac{d^2k_\perp}{(2\pi)^2}\Bigg[\ln\Big[&(1-r^H_\sigma r^H_\sigma e^{-2\kappa a})(1-r^H_\sigma r^H_\sigma e^{-2\kappa a})(1-r^H_\sigma r^H_\sigma e^{-2\kappa a})\\ &-(1-r^H_\sigma r^H_\sigma e^{-2\kappa a})(r^H_\sigma (t^H_\sigma)^2 r^H_\sigma e^{-4\kappa a}) - (r^H_\sigma (t^H_\sigma)^2 r^H_\sigma e^{-4\kappa a})(1-r^H_\sigma r^H_\sigma e^{-2\kappa a}) - (r^H_\sigma (t^H_\sigma)^2 (t^H_\sigma)^2 r^H_\sigma e^{-6\kappa a})\Big]\\ &+\ln\Big[(1-r^E_\sigma r^E_\sigma e^{-2\kappa a})(1-r^E_\sigma r^E_\sigma e^{-2\kappa a})(1-r^E_\sigma r^E_\sigma e^{-2\kappa a})\\ &-(1-r^E_\sigma r^E_\sigma e^{-2\kappa a})(r^E_\sigma (t^E_\sigma)^2 r^E_\sigma e^{-4\kappa a}) - (r^E_\sigma (t^E_\sigma)^2 r^E_\sigma e^{-4\kappa a})(1-r^E_\sigma r^E_\sigma e^{-2\kappa a}) - (r^E_\sigma (t^E_\sigma)^2 (t^E_\sigma)^2 r^E_\sigma e^{-6\kappa a})\Big]\Bigg]\end{aligned} \tag{27}$$

from considering $r_{1,2,3,4} = r_\sigma$, $t_{1,2,3,4} = t_\sigma$, $r_{5,6} = 0$ and constant $l_{ij} = a$ in Eq. (7). Introducing the optical coefficients of $\sigma$ conductivity $\delta$-plate $r_\sigma, t_\sigma$ from Eq. (11), expanding inside the logarithm, scaling with $\Delta E^c_{(12)}$, introducing the spherical polar coordinates and evaluation of $s$ integral gives

$$\frac{\Delta E_{(1234)}}{\Delta E^c_{(12)}} = \frac{45}{\pi^4}\int_0^1 dt\left[\sum_{i=1}^{3}\mathrm{Li}_4\left(\frac{1}{\alpha_i}\right)+\sum_{i=1}^{3}\mathrm{Li}_4\left(\frac{1}{\beta_i}\right)\right], \tag{28}$$

where $\alpha_1, \alpha_2, \alpha_3$ are the roots of the polynomial equation $1 + a_1x + b_1x^2 + c_1x^3 = 0$, and $\beta_1, \beta_2, \beta_3$ be the roots of $1 + a_2x + b_2x^2 + c_2x^3 = 0$ with

$$a_1 = -3\left(\frac{\sigma}{\sigma+2t}\right)^2,\ a_2 = -3\left(-\frac{\sigma}{\sigma+\frac{2}{t}}\right)^2, \tag{29}$$

are coefficients of $e^{-2s}$,

$$b_1 = 3\left(\frac{\sigma}{\sigma+2t}\right)^4 - 2\left(\frac{\sigma}{\sigma+2t}\right)^2\left(1-\frac{\sigma}{\sigma+2t}\right)^2,\ b_2 = 3\left(-\frac{\sigma}{\sigma+\frac{2}{t}}\right)^4 - 2\left(-\frac{\sigma}{\sigma+\frac{2}{t}}\right)^2\left(1-\frac{\sigma}{\sigma+\frac{2}{t}}\right)^2 \tag{30}$$

are coefficients of $e^{-4s}$ and

$$c_1 = -\left(\left(\frac{\sigma}{\sigma+2t}\right)^3 - \left(\frac{\sigma}{\sigma+2t}\right)\left(1-\frac{\sigma}{\sigma+2t}\right)^2\right)^2,\ c_2 = -\left(\left(-\frac{\sigma}{\sigma+\frac{2}{t}}\right)^3 - \left(-\frac{\sigma}{\sigma+\frac{2}{t}}\right)\left(1-\frac{\sigma}{\sigma+\frac{2}{t}}\right)^2\right)^2 \tag{31}$$

are coefficients of $e^{-6s}$.

## ACKNOWLEDGEMENTS

We are incredibly grateful to Dr. K. V. Shajesh and Dr. Prachi Parashar for introducing their work on $\delta$-function plates and for further guidance.

Data Availability Statement: No data associated with the manuscript.

---

[1] H. B. G. Casimir, "On the Attraction Between Two Perfectly Conducting Plates," *Kon. Ned. Akad. Wetensch. Proc.*, vol. 51, p. 793, 1948.
[2] E. M. Lifshitz, "The theory of molecular attractive forces between solids," *Sov. Phys. JETP*, vol. 2, pp. 73–83, 1956.

[3] B. W. Ninham and V. A. Parsegian, “van der Waals Interactions in Multilayer Systems,” *The Journal of Chemical Physics*, vol. 53, pp. 3398–3402, 11 1970.
[4] B. W. Ninham and V. A. Parsegian, “van der Waals Forces across Triple-Layer Films,” *The Journal of Chemical Physics*, vol. 52, pp. 4578–4587, 05 1970.
[5] E. Buks and M. L. Roukes, “Stiction, adhesion energy, and the casimir effect in micromechanical systems,” *Phys. Rev. B*, vol. 63, p. 033402, Jan 2001.
[6] H. B. Chan, V. A. Aksyuk, R. N. Kleiman, D. J. Bishop, and F. Capasso, “Nonlinear micromechanical casimir oscillator,” *Phys. Rev. Lett.*, vol. 87, p. 211801, Oct 2001.
[7] R. Balian and B. Duplantier, “Electromagnetic waves near perfect conductors. I. Multiple scattering expansions. distribution of modes,” *Annals of Physics*, vol. 104, no. 2, pp. 300–335, 1977.
[8] R. Balian and B. Duplantier, “Electromagnetic waves near perfect conductors. II. Casimir effect,” *Annals of Physics*, vol. 112, no. 1, pp. 165–208, 1978.
[9] O. Kenneth and I. Klich, “Opposites attract: A theorem about the Casimir force,” *Phys. Rev. Lett.*, vol. 97, p. 160401, Oct 2006.
[10] T. Emig, N. Graham, R. L. Jaffe, and M. Kardar, “Casimir forces between arbitrary compact objects,” *Phys. Rev. Lett.*, vol. 99, p. 170403, Oct 2007.
[11] K. S. Novoselov, A. K. Geim, S. V. Morozov, D. Jiang, Y. Zhang, S. V. Dubonos, I. V. Grigorieva, and A. A. Firsov, “Electric field effect in atomically thin carbon films,” *Science*, vol. 306, no. 5696, pp. 666–669, 2004.
[12] A. H. Castro Neto, F. Guinea, N. M. R. Peres, K. S. Novoselov, and A. K. Geim, “The electronic properties of graphene,” *Rev. Mod. Phys.*, vol. 81, pp. 109–162, Jan 2009.
[13] J. F. Dobson, A. White, and A. Rubio, “Asymptotics of the dispersion interaction: Analytic benchmarks for van der Waals energy functionals,” *Phys. Rev. Lett.*, vol. 96, p. 073201, Feb 2006.
[14] G. Gómez-Santos, “Thermal van der Waals interaction between graphene layers,” *Phys. Rev. B*, vol. 80, p. 245424, Dec 2009.
[15] M. Bordag, I. V. Fialkovsky, D. M. Gitman, and D. V. Vassilevich, “Casimir interaction between a perfect conductor and graphene described by the dirac model,” *Phys. Rev. B*, vol. 80, p. 245406, Dec 2009.
[16] D. Drosdoff and L. M. Woods, “Casimir forces and graphene sheets,” *Phys. Rev. B*, vol. 82, p. 155459, Oct 2010.
[17] J. S. Bunch, A. M. van der Zande, S. S. Verbridge, I. W. Frank, D. M. Tanenbaum, J. M. Parpia, H. G. Craighead, and P. L. McEuen, “Electromechanical resonators from graphene sheets,” *Science*, vol. 315, no. 5811, pp. 490–493, 2007.
[18] N. Inui, “Casimir effect on graphene resonator,” *Journal of Applied Physics*, vol. 119, p. 104502, 03 2016.
[19] E. M. Chudnovsky and R. Zarzuela, “Stability of suspended graphene under casimir force,” *Phys. Rev. B*, vol. 94, p. 085424, Aug 2016.
[20] M. S. Tomaš, “Casimir force in absorbing multilayers,” *Phys. Rev. A*, vol. 66, p. 052103, 2002.
[21] M. S. Tomaš, “Casimir effect across a layered medium,” *International Journal of Modern Physics: Conference Series*, vol. 14, pp. 561–565, 2012.
[22] N. Emelianova, N. Khusnutdinov, and R. Kashapov, “Casimir effect for a stack of graphene sheets,” *Phys. Rev. B*, vol. 107, p. 235405, Jun 2023.
[23] N. Khusnutdinov, R. Kashapov, and L. M. Woods, “Casimir effect for a stack of conductive planes,” *Phys. Rev. D*, vol. 92, p. 045002, Aug 2015.
[24] L. P. Teo, “Casimir piston of real materials and its application to multilayer models,” *Phys. Rev. A*, vol. 81, p. 032502, Mar 2010.
[25] P. S. Davids, F. Intravaia, F. S. S. Rosa, and D. A. R. Dalvit, “Modal approach to casimir forces in periodic structures,” *Phys. Rev. A*, vol. 82, p. 062111, Dec 2010.
[26] E. Amooghorban, M. Wubs, N. A. Mortensen, and F. Kheirandish, “Casimir forces in multilayer magnetodielectrics with both gain and loss,” *Phys. Rev. A*, vol. 84, p. 013806, Jul 2011.
[27] A. Allocca, S. Avino, S. Balestrieri, E. Calloni, S. Caprara, M. Carpinelli, L. D’Onofrio, D. D’Urso, R. De Rosa, L. Errico, G. Gagliardi, M. Grilli, V. Mangano, M. Marsella, L. Naticchioni, A. Pasqualetti, G. P. Pepe, M. Perciballi, L. Pesenti, P. Puppo, P. Rapagnani, F. Ricci, L. Rosa, C. Rovelli, D. Rozza, P. Ruggi, N. Saini, V. Sequino, V. Sipala, D. Stornaiuolo, F. Tafuri, A. Tagliacozzo, I. Tosta e Melo, and L. Trozzo, “Casimir energy for N superconducting cavities: a model for the YBCO (GdBCO) sample to be used in the Archimedes experiment,” *The European Physical Journal Plus*, vol. 137, no. 7, p. 826, 2022.
[28] J. Fiedler, K. Berland, J. W. Borchert, R. W. Corkery, A. Eisfeld, D. Gelbwaser-Klimovsky, M. M. Greve, B. Holst, K. Jacobs, M. Krüger, D. F. Parsons, C. Persson, M. Presselt, T. Reisinger, S. Scheel, F. Stienkemeier, M. Tømterud, M. Walter, R. T. Weitz, and J. Zalieckas, “Perspectives on weak interactions in complex materials at different length scales,” *Phys. Chem. Chem. Phys.*, vol. 25, pp. 2671–2705, 2023.
[29] T. Gong, M. R. Corrado, A. R. Mahbub, C. Shelden, and J. N. Munday, “Recent progress in engineering the casimir effect–applications to nanophotonics, nanomechanics, and chemistry,” *Nanophotonics*, vol. 10, no. 1, pp. 523–536, 2021.
[30] A. Stange, D. K. Campbell, and D. J. Bishop, “Science and technology of the Casimir effect,” *Physics Today*, vol. 74, pp. 42–48, 01 2021.
[31] A. A. Geraci, S. B. Papp, and J. Kitching, “Short-range force detection using optically cooled levitated microspheres,” *Phys. Rev. Lett.*, vol. 105, p. 101101, Aug 2010.
[32] T. Westphal, H. Hepach, J. Pfaff, and M. Aspelmeyer, “Measurement of gravitational coupling between millimetre-sized masses,” *Nature*, vol. 591, pp. 225–228, Mar 2021.

[33] C. P. Blakemore, A. Fieguth, A. Kawasaki, N. Priel, D. Martin, A. D. Rider, Q. Wang, and G. Gratta, "Search for non-newtonian interactions at micrometer scale with a levitated test mass," *Phys. Rev. D*, vol. 104, p. L061101, Sep 2021.

[34] Y.-J. Chen, W. K. Tham, D. E. Krause, D. López, E. Fischbach, and R. S. Decca, "Stronger limits on hypothetical yukawa interactions in the 30–8000 nm range," *Phys. Rev. Lett.*, vol. 116, p. 221102, Jun 2016.

[35] V. Abhignan, "Casimir energy of $N$ magnetodielectric $\delta$-function plates," *Physica Scripta*, vol. 98, p. 105018, sep 2023.

[36] V. Abhignan, "Quasiperiodic arrangement of magnetodielectric -plates: Green's functions and casimir energies for n bodies," *Physica Scripta*, vol. 100, p. 045006, mar 2025.

[37] P. Parashar, K. A. Milton, K. V. Shajesh, and M. Schaden, "Electromagnetic semitransparent $\delta$-function plate: Casimir interaction energy between parallel infinitesimally thin plates," *Phys. Rev. D*, vol. 86, p. 085021, Oct 2012.

[38] G. Barton, "Casimir effects for a flat plasma sheet: I. energies," *Journal of Physics A: Mathematical and General*, vol. 38, no. 13, pp. 2997–3019, 2005.

[39] G. Barton, "Casimir effects for a flat plasma sheet: Ii. fields and stresses," *Journal of Physics A: Mathematical and General*, vol. 38, p. 3021, mar 2005.

[40] L. A. Falkovsky and A. A. Varlamov, "Space-time dispersion of graphene conductivity," *The European Physical Journal B*, vol. 56, pp. 281–284, Apr 2007.

[41] I. V. Fialkovsky and D. V. Vassilevich, "Graphene through the looking glass of QFT," *Modern Physics Letters A*, vol. 31, no. 40, p. 1630047, 2016.

[42] A. B. Kuzmenko, E. van Heumen, F. Carbone, and D. van der Marel, "Universal optical conductance of graphite," *Phys. Rev. Lett.*, vol. 100, p. 117401, Mar 2008.

[43] R. R. Nair, P. Blake, A. N. Grigorenko, K. S. Novoselov, T. J. Booth, T. Stauber, N. M. R. Peres, and A. K. Geim, "Fine structure constant defines visual transparency of graphene," *Science*, vol. 320, no. 5881, pp. 1308–1308, 2008.

[44] D. Drosdoff, A. D. Phan, L. M. Woods, I. V. Bondarev, and J. F. Dobson, "Effects of spatial dispersion on the casimir force between graphene sheets," *The European Physical Journal B*, vol. 85, p. 365, Nov 2012.

[45] G. Carleo, I. Cirac, K. Cranmer, L. Daudet, M. Schuld, N. Tishby, L. Vogt-Maranto, and L. Zdeborová, "Machine learning and the physical sciences," *Rev. Mod. Phys.*, vol. 91, p. 045002, Dec 2019.

[46] G. Karagiorgi, G. Kasieczka, S. Kravitz, B. Nachman, and D. Shih, "Machine learning in the search for new fundamental physics," *Nature Reviews Physics*, vol. 4, pp. 399–412, Jun 2022.

[47] M. N. Chernodub, H. Erbin, I. V. Grishmanovskii, V. A. Goy, and A. V. Molochkov, "Casimir effect with machine learning," *Phys. Rev. Res.*, vol. 2, p. 033375, Sep 2020.

[48] I. Brevik, P. Parashar, and K. V. Shajesh, "Casimir force for magnetodielectric media," *Phys. Rev. A*, vol. 98, p. 032509, Sep 2018.

[49] K. V. Shajesh, I. Brevik, I. Cavero-Peláez, and P. Parashar, "Casimir energies of self-similar plate configurations," *Phys. Rev. D*, vol. 94, p. 065003, Sep 2016.

[50] T. H. Boyer, "Van der Waals forces and zero-point energy for dielectric and permeable materials," *Phys. Rev. A*, vol. 9, pp. 2078–2084, May 1974.

[51] M. Asorey and J. Muñoz-Castañeda, "Attractive and repulsive Casimir vacuum energy with general boundary conditions," *Nuclear Physics B*, vol. 874, no. 3, pp. 852–876, 2013.

[52] MathWorks, "Matlab Deep Learning Toolbox," *https://www.mathworks.com/help/deeplearning/ref/fitnet.html*, 2024.

[53] A. B. Djurišić and E. H. Li, "Optical properties of graphite," *Journal of Applied Physics*, vol. 85, pp. 7404–7410, 05 1999.

[54] Y. Yao, M. A. Kats, P. Genevet, N. Yu, Y. Song, J. Kong, and F. Capasso, "Broad electrical tuning of graphene-loaded plasmonic antennas," *Nano Letters*, vol. 13, pp. 1257–1264, Mar 2013.

[55] F. J. García de Abajo, "Graphene plasmonics: Challenges and opportunities," *ACS Photonics*, vol. 1, pp. 135–152, Mar 2014.

[56] T. Low and P. Avouris, "Graphene plasmonics for terahertz to mid-infrared applications," *ACS Nano*, vol. 8, pp. 1086–1101, Feb 2014.

[57] E. H. Hwang and S. Das Sarma, "Dielectric function, screening, and plasmons in two-dimensional graphene," *Phys. Rev. B*, vol. 75, p. 205418, May 2007.

[58] B. Wunsch, T. Stauber, F. Sols, and F. Guinea, "Dynamical polarization of graphene at finite doping," *New Journal of Physics*, vol. 8, p. 318, dec 2006.

[59] G. Lovat, G. W. Hanson, R. Araneo, and P. Burghignoli, "Semiclassical spatially dispersive intraband conductivity tensor and quantum capacitance of graphene," *Phys. Rev. B*, vol. 87, p. 115429, Mar 2013.

[60] P. Rodriguez-Lopez, W. J. M. Kort-Kamp, D. A. R. Dalvit, and L. M. Woods, "Nonlocal optical response in topological phase transitions in the graphene family," *Phys. Rev. Mater.*, vol. 2, p. 014003, Jan 2018.

[61] K. Bukvišová, R. Kalousek, M. Patočka, J. Zlámal, J. Planer, V. Mahel, D. Citterberg, L. Novák, T. Šikola, S. Kodambaka, and M. Kolíbal, "Casimir-like effect driven self-assembly of graphene on molten metals," 2025.

[62] S. Das, J. Fiedler, O. Stauffert, M. Walter, S. Y. Buhmann, and M. Presselt, "Macroscopic quantum electrodynamics and density functional theory approaches to dispersion interactions between fullerenes," *Phys. Chem. Chem. Phys.*, vol. 22, pp. 23295–23306, 2020.

[63] M. S. Okyay, M. Choi, Q. Xu, A. P. Diéguez, M. Del Ben, K. Z. Ibrahim, and B. M. Wong, "Unconventional nonlinear hall effects in twisted multilayer 2d materials," *npj 2D Materials and Applications*, vol. 9, p. 1, Jan 2025.

[64] I. Dzyaloshinskii, E. Lifshitz, and L. Pitaevskii, "The general theory of Van der Waals forces," *Advances in Physics*, vol. 10, no. 38, pp. 165–209, 1961.

[65] A. A. Feiler, L. Bergström, and M. W. Rutland, "Superlubricity using repulsive van der waals forces," *Langmuir*, vol. 24, pp. 2274–2276, Mar 2008.
[66] J. N. Munday, F. Capasso, and V. A. Parsegian, "Measured long-range repulsive Casimir–Lifshitz forces," *Nature*, vol. 457, no. 7226, pp. 170–173, 2009.
[67] O. Kenneth, I. Klich, A. Mann, and M. Revzen, "Repulsive Casimir forces," *Phys. Rev. Lett.*, vol. 89, p. 033001, Jun 2002.
[68] M. Levin, A. P. McCauley, A. W. Rodriguez, M. T. H. Reid, and S. G. Johnson, "Casimir repulsion between metallic objects in vacuum," *Phys. Rev. Lett.*, vol. 105, p. 090403, Aug 2010.
[69] Y. Zhang, H. Zhang, X. Wang, Y. Wang, Y. Liu, S. Li, T. Zhang, C. Fan, and C. Zeng, "Magnetic-field tuning of the casimir force," *Nature Physics*, vol. 20, pp. 1282–1287, Aug 2024.
[70] D. Iannuzzi and F. Capasso, "Comment on "repulsive Casimir forces"," *Phys. Rev. Lett.*, vol. 91, p. 029101, Jul 2003.
[71] O. Kenneth, I. Klich, A. Mann, and M. Revzen, "Kenneth et al. reply:," *Phys. Rev. Lett.*, vol. 91, p. 029102, Jul 2003.
[72] V. M. Shalaev, "Optical negative-index metamaterials," *Nature Photonics*, vol. 1, pp. 41–48, Jan 2007.
[73] I. Pirozhenko and A. Lambrecht, "Repulsive Casimir forces and the role of surface modes," *Phys. Rev. A*, vol. 80, p. 042510, Oct 2009.
[74] V. Yannopapas and N. V. Vitanov, "First-principles study of Casimir repulsion in metamaterials," *Phys. Rev. Lett.*, vol. 103, p. 120401, Sep 2009.
[75] F. S. S. Rosa, D. A. R. Dalvit, and P. W. Milonni, "Casimir interactions for anisotropic magnetodielectric metamaterials," *Phys. Rev. A*, vol. 78, p. 032117, Sep 2008.
[76] F. S. S. Rosa, D. A. R. Dalvit, and P. W. Milonni, "Casimir-lifshitz theory and metamaterials," *Phys. Rev. Lett.*, vol. 100, p. 183602, May 2008.
[77] D. Drosdoff and L. M. Woods, "Casimir interactions between graphene sheets and metamaterials," *Phys. Rev. A*, vol. 84, p. 062501, Dec 2011.
[78] K. S. Burch, D. Mandrus, and J.-G. Park, "Magnetism in two-dimensional van der waals materials," *Nature*, vol. 563, pp. 47–52, Nov 2018.
[79] C. Gong and X. Zhang, "Two-dimensional magnetic crystals and emergent heterostructure devices," *Science*, vol. 363, no. 6428, p. eaav4450, 2019.
[80] C. Shelden, B. Spreng, and J. N. Munday, "Enhanced repulsive casimir forces between gold and thin magnetodielectric plates," *Phys. Rev. A*, vol. 108, p. 032817, Sep 2023.
[81] J. S. Høye and I. Brevik, "Repulsive Casimir force," *Phys. Rev. A*, vol. 98, p. 022503, Aug 2018.